\documentclass{article}

\usepackage{emulateapj1}

\usepackage{psfig}
\usepackage{times}

\def\ledd{L_{\rm Edd}}
\def\luv{L_{\rm uv}}

\newcommand\ltot{L_{\rm tot}}

\newcommand\rozanska{R\'o$\dot{\rm z}$a\'nska }
\newcommand\zycki{$\dot{\rm Z}$ycki }

\newcommand\fx{F_{\rm x}}

\newcommand\fdisk{F_{\rm d}}

\newcommand\pg{P_{\rm gas}}

\newcommand\teff{T_{\rm eff}}

\newcommand\tskin{\tau_{\rm s}}

\newcommand\dm{\dot{m}}
\newcommand\ztop{z_{\rm t}}

\def\>{$>$}
\def\<{$<$}

\def\simlt{\lower.5ex\hbox{$\; \buildrel < \over \sim \;$}}
\def\simgt{\lower.5ex\hbox{$\; \buildrel > \over \sim \;$}}
\def\sqr#1#2{{\vcenter{\hrule height.#2pt
      \hbox{\vrule width.#2pt height#1pt \kern#1pt
         \vrule width.#2pt}
      \hrule height.#2pt}}}

\begin{document}


\title{Accretion disk models and their X-ray reflection
signatures. I. Local spectra.}

\author{Sergei Nayakshin\altaffilmark{1} and Timothy R. Kallman}

\affil{NASA/GSFC, LHEA, Code 661, Greenbelt, MD, 20771}
\altaffiltext{1}{National Research Council Associate;
email: serg@milkyway.gsfc.nasa.gov}

\begin{abstract}
X-ray illumination of accretion disks is an invaluable diagnostic of
the structure of these disks because of the associated iron K$\alpha$
emission. Here we point out that the resulting reflected spectra
depend very sensitively on the geometry of the X-ray source, and that
this fact can be efficiently used to test these models
observationally. In particular, we discuss three different accretion
disk geometries: the ``lamppost model'', accretion disks with magnetic
flares, and the model with a full corona overlying a cold thin
disk. We show that in the case of the lamppost model, unless the X-ray
luminosity of the central source is larger than that of the cold disk
by a factor of 10 or more, a significant fraction of iron in the
ionized skin of the disk is in the hydrogen and helium-like ions.
Because these ions have large fluorescence yields, the resulting
reflected spectra look strongly ionized, with Equivalent Width (EW) of
the line {\em increasing} with X-ray luminosity $L_x$ up to the
maximum of $\sim 500$ eV. This situation contrasts to the magnetic
flare model, where the large X-ray flux near flares completely ionizes
the skin of the disk and thus the resulting spectra appear to be that
from a neutral material. The line EW in this model {\em
anti-correlates} with X-ray luminosity, and becomes arbitrarily small
when $L_x$ is a good fraction of the Eddington luminosity.  Finally,
in the full corona case, due to the additional pressure and weight of
the corona, the gas pressure (and its density) below the corona is
always large enough to make the gas very cool and effectively
neutral. No highly ionized skin forms in such a model.  If the corona
is Thomson thin, then EW of the line does not depend on the accretion
disk or corona luminosities for the full corona model.
\end{abstract}


\section{Introduction}\label{sect:intro}

X-ray illumination of accretion disks in Active Galactic Nuclei (AGN)
and Galactic Black Hole Candidates (GBHC) is a phenomenon of a great
observational importance with implications for theories of accretion
disks (AD). Since X-rays often produce a non-trivial part of the
overall bolometric luminosity of AGN and GBHCs, it is clear that X-ray
heating of the accretion disk surface may change the energy and
ionization balance there, causing corresponding changes across the
entire electromagnetic spectrum emitted by these objects. In addition,
detailed calculations predict that spectra should contain many
potentially observable atomic lines, edges, recombination continua,
etc. Therefore, comparison of theoretical models (which are
parameterized by few parameters only) and observations presents an
invaluable opportunity to solve the inverse problem of Astrophysics of
ADs -- that is to learn about the accretion disk structure from
observed spectra.

This is why so much theoretical effort has gone into studies of the
X-ray illumination problem in the last few decades. A first detailed
account of the problem can be found in Basko, Sunyaev \& Titarchuck
(1974)\footnote{These authors considered X-ray illumination of the
surface of a normal star in an X-ray binary, but this problem is
almost identical to the one of X-ray illuminated ADs.}. A further
important development was done by Lightman \& White (1988) and White,
Lightman \& Zdziarski (1988) who provided fitting formulae to the
results of their Monte-Carlo simulations of reflection off {\em
neutral} matter, which allowed X-ray observers to use these formulae
to make direct fits to data. In addition, Fabian et al. (1989) have
shown that the relativistically smeared fluorescent K$\alpha$ line
emission of ADs yields characteritic line profiles that can be used to
constrain geometry of accretion disks.  A number of authors expanded
on these studies since then (e.g., George \& Fabian 1991; Done et
al. 1992; Ross \& Fabian 1993; Matt, Brandt \& Fabian 1996; \zycki et
al. 1994; Czerny \& \zycki 1994; Krolik, Madau \& \zycki 1994;
Magdziarz \& Zdziarski 1995; Ross, Fabian \& Brandt 1996; Matt, Fabian
\& Ross 1993, 1996; Poutanen, Nagendra \& Svensson 1996; Blackman
1999)

However, except for Basko et al. (1974), the authors of the
publications referenced above either studied a non-ionized reflection,
or assumed that the density of the illuminated gas is constant with
height for the more complicated ionized reflection calculations.
Raymond (1993) and Ko \& Kallman (1994) considered the problem of
X-ray illumination of an accretion disk in Low Mass X-ray Binaries
(LMXB) relaxing the assumption of the constant gas density and instead
solving for the density via hydrostatic balance. \rozanska \& Czerny
(1996) also included hydrostatic balance in their semi-analytical
study of X-ray illumination in AGN. Nayakshin, Kazanas \& Kallman
(2000; hereafter NKK) extended results of \rozanska \& Czerny (1996)
for the inner parts of ADs in AGN and GBHCs by providing an accurate
radiation transfer in the optically thick illuminated slab.  All these
authors found that the thermal ionization instability, previously well
known in the context of the AGN emission line regions (e.g., Krolik,
McKee \& Tarter 1981), plays a crucial role in the establishing of the
equilibrium temperature and density profiles of the X-ray illuminated
gas. Through these profiles, the instability is directly involved in
the formation of the reflected continuum spectrum as well as
fluorescent line emission, such as that of iron K$\alpha$ lines.

While the fixed density models are out of hydrostatic balance (e.g.,
the gas pressure at the top of the illuminated layer may be up to few
hundred times larger than that on the bottom of the layer), one could
still hope that the main results of such calculations will apply to
the more realistic hydrostatic balance models. However, NKK compared
spectra obtained from these two classes of models, and found that the
predictions of these two models for the behavior of the iron K$\alpha$
lines, edges, and all the other features of the reflected spectrum are
very different.

In this paper, we use the code of NKK to make a detailed analysis of
local reflected spectra in two physically distinct limits\footnote{We
specifically limit our attention to single radius unsmeared spectra to
expose the physics of the problem clearer. Full disk spectra with
relativistic energy shifts will be presented in a separate
publication}. We will show that these two classes of models lead to
completely different reflection spectra unless accretion rate is very
small ($\dm\simlt 10^{-3}$, see below).

In one limit, the X-ray flux illuminating the disk, $\fx$, is smaller
than $\fdisk$, the flux of the soft thermal emission intrinsically
generated in the disk.  Physically, the situation $\fx\simlt \fdisk$
occurs in the so-called ``lamppost model'', where the X-rays are
produced high above the black hole, so that they illuminate a large
portion of the innermost disk region. Roughly speaking, the X-ray
illuminating flux is $\fx \sim L_x/4\pi R^2$, where $R \sim 10 R_S$,
where $R_s$ is Schwartzchild radius.

In the other case, the X-ray flux exceeds the disk flux by a large
factor. This situation occurs when the X-ray luminosity is produced
within magnetic flares, such that most of the X-ray reflection happens
near the flare locations. The physical distinction from the lamppost
model is that the same X-ray luminosity originates much closer to the
disk surface because magnetic loops are expected to be of the order of
few disk height scales, which is much smaller than radius for thin
disks. Hence, the X-ray illumination will be spread over the disk
surface very unevenly.  The covering fraction of magnetic flares,
$f_c$, may be quite small (see estimates in Nayakshin 1998b, \S
2.5.5). Most of the disk will receive little X-ray illumination,
whereas near the flares $\fx \sim L_x/4\pi f_c R^2 \gg L_x/4\pi R^2$
and thus $\fx$ is very likely to exceed $\fdisk$. It is also well
known that $\fx\gg \fdisk$ is in fact a necessary condition for the
magnetic flare model to reproduce the continuum X-ray and UV spectra
of AGN and GBHCs (see Haardt, Maraschi \& Ghisellini 1994; Svensson
1996; Nayakshin 1998a).

In addition, we discuss the ionization equilibria in the geometry of a
full corona overlying a cold accretion disk. We consider coronae
heated from below and also coronae heated by internal viscous
dissipation (as those thought to exist in the transition region of the
Advection Dominated Accretion Flows -- e.g., Esin, McClintok \&
Narayan 1997).  In both of these cases we show that the ionization
equilibria permit only cold solutions for the material below the
corona. Therefore, reflected spectra from such a material should look
``neutral''.  We believe that the predictions of these three models
are sufficiently different to allow these models to be be meaningfully
tested against observations with existing and future data.

The structure of the paper is as follows. In \S 2 we describe the way
in which several parameters important for the X-ray illumination
caluclations can be deduced for any accretion disk theory once broad
band spectrum of an accreting source is known. In \S
\ref{sect:lampost} and \S \ref{sect:flares} we present our
calculations for the lamppost and magnetic flare models,
respectively. In \S \ref{sect:fullcorona} we consider the X-ray
illumination for the full corona case. We give an extended discussion
of our results in \S \ref{sect:discussion} and in \S
\ref{sect:conclusions} we present our conclusions.

\section{On predictive power of X-ray illumination calculations}
\label{sect:predict}

Calculations of the X-ray reflected spectra is a very powerful tool
with which to infer the structure of ADs around compact objects. The
value of such calculations is currently under-estimated, we believe,
and this is why we will now explain how these calculations should be
used in order to constrain accretion disk theories and why the results
of such calculations are quite robust. Let us assume that we have a
well resolved broad band spectrum of an AGN with the total luminosity
$\ltot$, and that the integrated optical-UV luminosity is $\luv$,
whereas that of the entire X-ray range is $L_x$ (so that $\ltot = \luv
+ L_x$).

One can now assume a value for the black hole mass, $M$, and then
investigate a particular accretion disk theory. For example, if the
disk structure is given by the Shakura-Sunyaev theory, then we can (1)
find the dimensionless accretion rate $\dm = \ltot/\ledd$, where
$\ledd$ is the Eddington luminosity for the mass $M$; (2) determine
the disk flux, $\fdisk$, and the disk height scale, $H$, for every
radius. The next step is to use the geometry of the X-ray source
appropriate for the given model to infer how the X-ray illuminating
flux is distributed over the disk surface. With this, {\em there are
no uncertainties in the resulting reflected spectra} beyond geometry
and value of $M$. This is because the gravity parameter $A$, defined
by NKK, is none other than the ratio of the vertical component of the
gravitational force, ${\cal F}_g$, at the height of one disk height
scale to the radiation pressure force, ${\cal F}_{\rm rad} \equiv
\sigma_t n_e \fx/c$, that the X-ray flux would provide if the cross
section were given by the Thomson value ($n_e = 1.2 n_H$ is the
electron density assuming that H and He are completely ionized):
\begin{equation}
A\equiv {R_s \mu_m H^2\over 2 R^3 m_e} \, {m_e c^3\over \sigma_t F_x
H}\, =
{GM\rho H\over R^3}\; {c\over \sigma_t n_e\fx} \equiv
{{\cal F}_g\over {\cal F}_{\rm rad}}\;,
\label{adef}
\end{equation}
Hence, once the accretion disk structure is prescribed, and $\fx(r)$
is known, the reflection calculations will provide a definite outcome
that can be compared with observations.

We emphasis that this has not been possible in the context of the
conventional constant density models.  In the context of those models,
the gas density in the midplane of the disk, $n_H$, is the primary
parameter, since it enters the definition of the ionization parameter
$\xi = 4\pi F_x/n_H$ (see, e.g., \zycki et al. 1994). Because
$n_H\propto \alpha^{-1}$ for radiation-dominated ADs, where $\alpha$
is the Shakura-Sunyaev viscosity parameter (see Shakura \& Sunyaev
1973), ionization parameter scales as $\xi\propto\alpha$. For
gas-dominated disks, recent MHD simulations give $\alpha\sim 0.01$
(e.g., Miller \& Stone 2000). Observations of disks in Cataclismic
Variables seem to imply $\alpha\sim 0.1$ (Smak 1984; 1999), but it is
also not unusual to invoke values of $\alpha = 0.3$ (e.g., Esin et
al. 1997) or even closer to unity. For radiation dominated disks, the
value of $\alpha$ is even less certain, because it is not clear
whether the viscosity in such disks will scale with total or only the
gas pressure (see, e.g., Stella \& Rosner 1984; Nayakshin, Rappaport
\& Melia 2000), so in principle the value of the ``efective'' $\alpha$
can be as small as $\sim 10^{-6}$. Therefore, the results of the
constant density models may be uncertain by a factor of $10^2- 10^6$
(!).

Our calculations avoid this problem because the value of the gas
density (or total pressure) in the midplane has very little influence
on the final result. The important parameter is the height, $z_b$, at
which the bottom of the ionized skin is located because that defines
the value of the gravitational force ${\cal F}_g$. The pressure at
$z_b$ is very small compared with the disk mid-plane pressure. Since
at $z\sim z_b$ the pressure declines very quickly with height
(exponentially -- see, e.g., Shakura \& Sunyaev 1973), a large change
in $\alpha$ will lead only to a logariphmic change in $z_b$ (see also
Nayakshin 2000 on that).  In other words, the most important parameter
of our calculations -- the gravity parameter $A$ -- depends on
$\alpha$ only logariphmically\footnote{This is similar to stellar
atmospheres -- clearly, it is the gravity and the star's radius that
are important, not the gas density in the center of the star.}.

Finally, one does not have to assume that the structure of the
illuminated AD is given by the Shakura-Sunyaev theory. Any other AD
theory may be used to prescribe the vertical disk structure and then
calculations proceed in exactly the same way as they would for a SS
disk.

\section{Lamp Post model: the ``warm skin'' limit.}\label{sect:lampost}

\subsection{Physical setup and method of
calculation}\label{sect:setup}

In this section we will assume that all the X-rays are produced within
a point-like source located $h_x = 6 R_s$ above the black hole on the
symmetry axis. Although there is no solid physical justification for
the location of the X-ray source directly above the black hole, this
geometry can be used as a testing ground for studying complex
phenomena, such as iron line reverberation (e.g., Reynolds \& Begelman
1997; Reynolds et al. 1999; Young \& Reynolds 2000).  We will neglect
special and general relativistic effects as we concentrate on the
physics of the local (i.e., at a given disk radius $R$) ionization
balance. We restrict our attention here to $R= 6 R_s$.

We use the code of NKK to solve the X-ray illumination
problem. Several input parameters are (1) the disk accretion rate,
$\dm$, measured in the Eddington units such that $\dm = 1$ corresponds
to the disk luminosity equal to the Eddington luminosity; (2) the
luminosity of the X-ray source, parameterized in terms of the ratio of
the latter to the disk integrated luminosity, $\eta_x$ (see Nayakshin
2000); (3) the cosine of the X-ray incidence angle that is fixed by
values of $h_x$ and $R$ ($\mu_i = 1/\sqrt{2}$). For the most tests in
this paper, we fixed $\eta_x$ at a value of $0.2$.

\subsection{Temperature profiles}\label{sect:tempw}

\begin{figure*}[T]
\centerline{\psfig{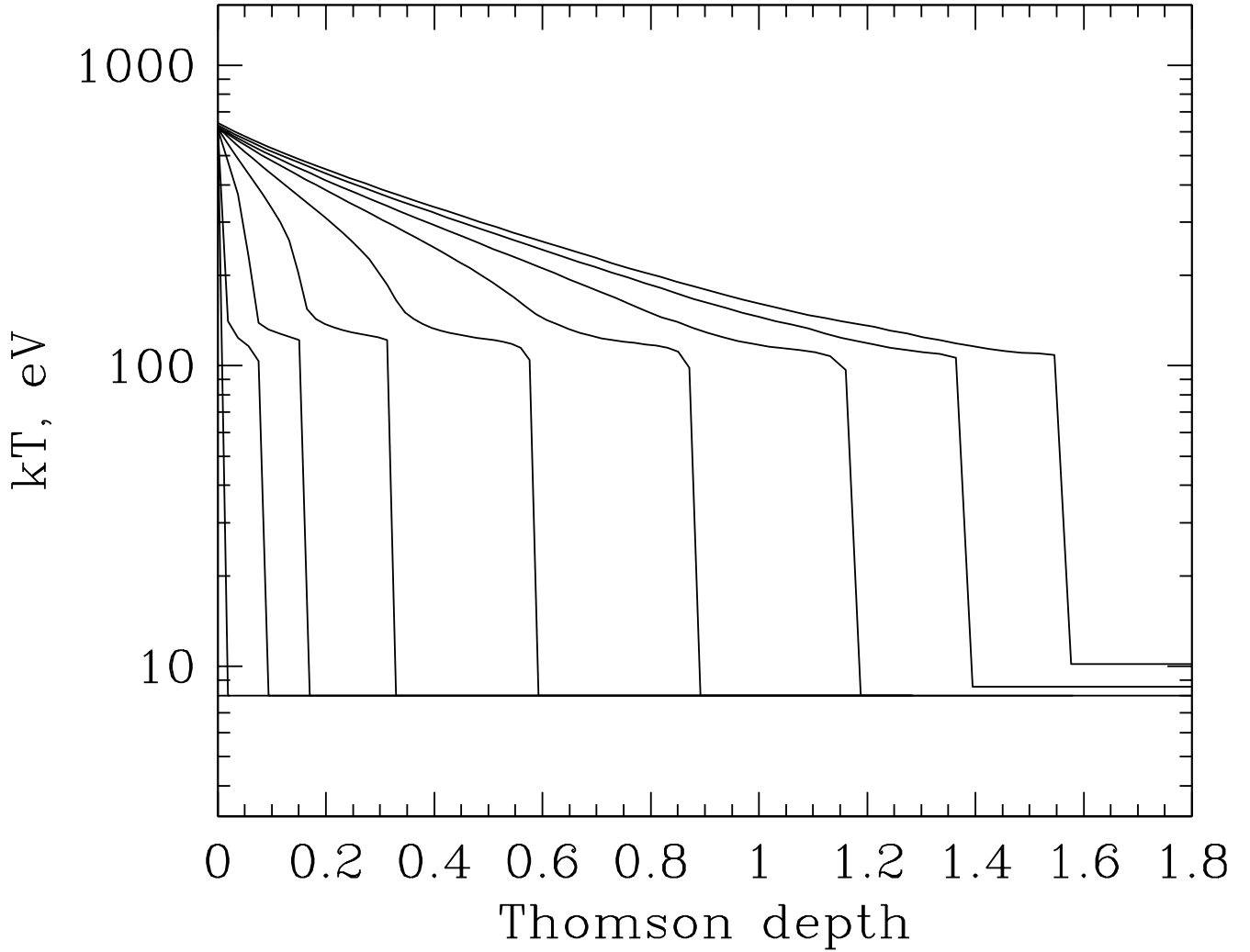}}
\caption{Gas temperature as a function of Thomson depth for the ``lamp
post'' model for accretion rates covering a range from $\dm = 10^{-3}$
to $0.512$. Starting from the second least ionized curve, the
accretion rate is increasing by a factor of 2 for each consequtive
curve. Note that the maximum gas temperature is far below the X-ray
Compton temperature $k T_x\sim 7$ keV because of the very large disk
intrinsic flux. This fact is the main reason why the spectrum looks
``highly ionized'' (see Fig. \ref{fig:spssh} below).}
\label{fig:temps}
\end{figure*}

Figure \ref{fig:temps} shows the resulting temperature profiles for
the X-ray illuminated upper layer of the disk for $\eta_x = 0.2$ and
the accretion rate $\dm$ scanning a range of values from the low value
of $10^{-3}$ to high of $0.512$. The case with $\dm = 10^{-3}$ is the
least ionized one, and it corresponds to the left most curve in
Fig. \ref{fig:temps} (only the first zone is highly ionized, and all
the rest are at the temperature $k T\simeq 8$ eV). The next curve to
the right was computed for $\dm= 4\times 10^{-3}$, and all the
subsequent curves are computed with $\dm$ increasing by a factor of 2
from a previous value. We will refer to these runs by their number,
such that the least ionized case is referred to as $w1$ and the most
ionized as $w9$ ($w$ stands for ``warm'').

The temperature of the ionized skin will later be shown to be decisive
in establishing the nature of the reflected spectra.  It is important
to note that the maximum gas temperature is almost the same in all of
these cases, and that it is far below $T_x$, the Compton temperature
corresponding to the X-ray flux only. This is why we refer to the given
limit as the ``warm skin'' limit to distinguish it from the ``hot
skin'' limit studied in \S \ref{sect:flares} below.  In the Eddington
approximation, one can easily show that the value of the Compton
temperature at the top of the skin is approximately given by
\begin{equation}
T_c = {J_x T_x + T_{\rm bb} J_{\rm bb}\over J_x + J_{\rm
bb}}\simeq T_x \left [ 1 + 2\mu_i \; {\fx + \fdisk \over \fx}\right
]^{-1}\;.
\label{tc}
\end{equation}
From this equation it is apparent that the skin is much cooler than
the X-ray Compton temperature because of the presence of the large
intrinsic disk flux, $\fdisk$. Therefore, physically, the warm skin
limit corresponds to the case $\fdisk \gg \fx$, whereas the hot skin
results when $\fdisk\ll \fx$ (and the X-ray spectrum is hard such that
$T_x$ is high).

Note also that in all cases considered, except for the run $w1$, there
is a rather extended ``shelf'' with $kT\sim 100$ eV, whereas this
shelf is absent or weak in the hot skin limit (see
Fig. \ref{fig:temp1} below). We give an explanation for this fact in
\S \ref{sect:ie}.

\subsection{Reflected spectra}\label{sect:sps}

\begin{figure*}[T]
\centerline{\psfig{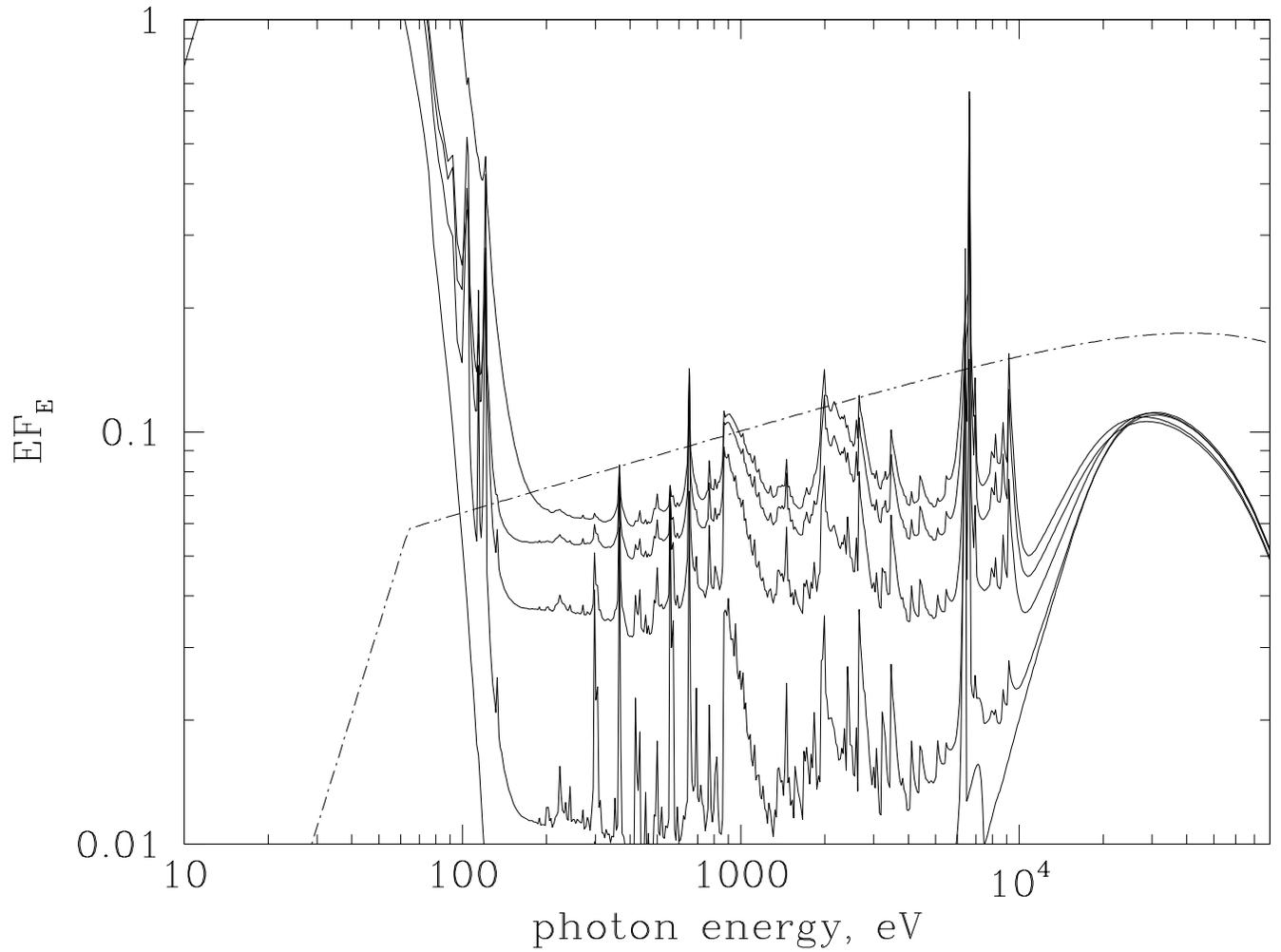}}
\caption{Angle-averaged reflected spectra for the lamp post model
(solid) for runs $w1$, $w3$, $w5$, $w7$ and $w9$. The spectra are
normalized such that the X-ray illuminated continuum flux is equal to
unity for all of them.}
\label{fig:sp_sall}
\end{figure*}

Figure \ref{fig:sp_sall} shows angle-averaged reflected spectra in the
range from 10 eV to about 100 keV. The ionizing X-ray spectrum is
drawn with the dashed-dotted curve. All the spectra are normalized in
such a way that $\fx^{-1}\,\int_0^{\infty} E\, dE\, F(E) = 1 +
\fdisk/\fx$ for convinience. The least ionized reflected spectrum
corresponds to the run $w1$, and the subsequently more ionized spectra
are for the tests $w3$, $w5$, $w7$, and $w9$. The coldest case ($w1$)
is basically the usual neutral reflection component that has almost no
reflected flux below few keV all the way down to $E\sim$ few $\times
k\teff$, where the emission is dominated by a quasi-black body
reprocessed flux.

\begin{figure*}[T]
\centerline{\psfig{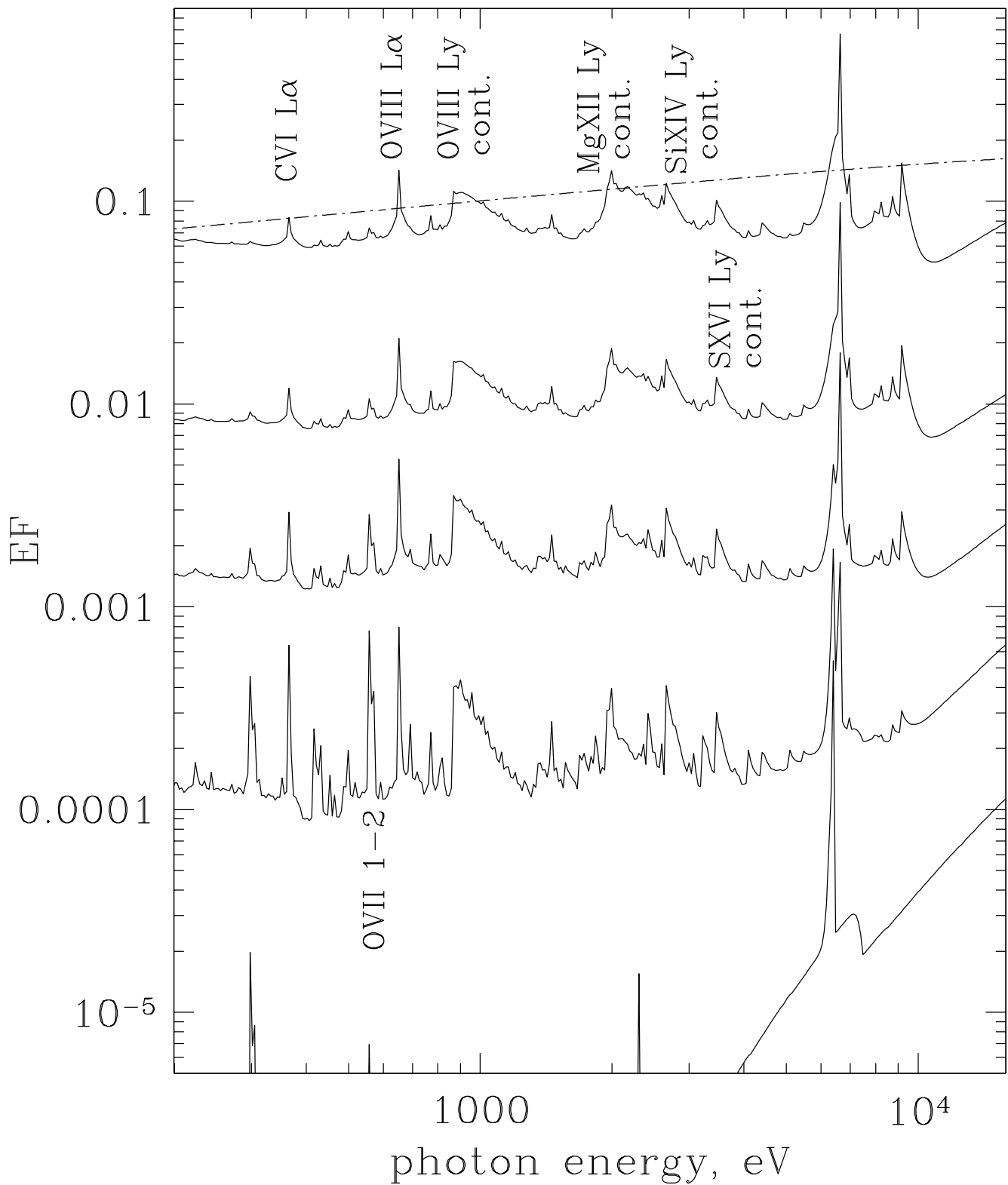}}
\caption{Same spectra as in Figure \ref{fig:sp_sall}, but shifted to
allow a better visibility of different atomic features. Note that
except for the least ionized test $w1$, all the spectra are distinctly
different from neutral reflection in the line emissivity, energy, iron
photo-absorption edge and many prominent features in soft X-rays.}
\label{fig:spssh}
\end{figure*}

Figure \ref{fig:spssh} shows the same spectra in a narrower energy
range. The spectra are also shifted with respect to each other to
allow for a greater visibility of the individual curves. It is
immediately clear that except for the least ionized case on the bottom
of the Figure, all the other spectra are strikingly different from the
neutral reflection spectrum. Note that the energy of the dominant line
component is around 6.7 keV and it comes from He-like iron mainly (see
\S \ref{sect:sion} below). The line Equivalen Width (EW) is about 180
eV in the coldest case and it increases to about 500 eV in the run
$w9$, which is consistent with earlier findings of Matt, Fabian \&
Ross (1996). Note also a very prominent ionized absorption edge and
the recombination continuum at $\sim 9$ keV. Other significant
spectral features are labeled directly in Figure \ref{fig:spssh}.

\subsection{Ionization Structure of the gas}\label{sect:sion}

\begin{figure*}[T]
\centerline{\psfig{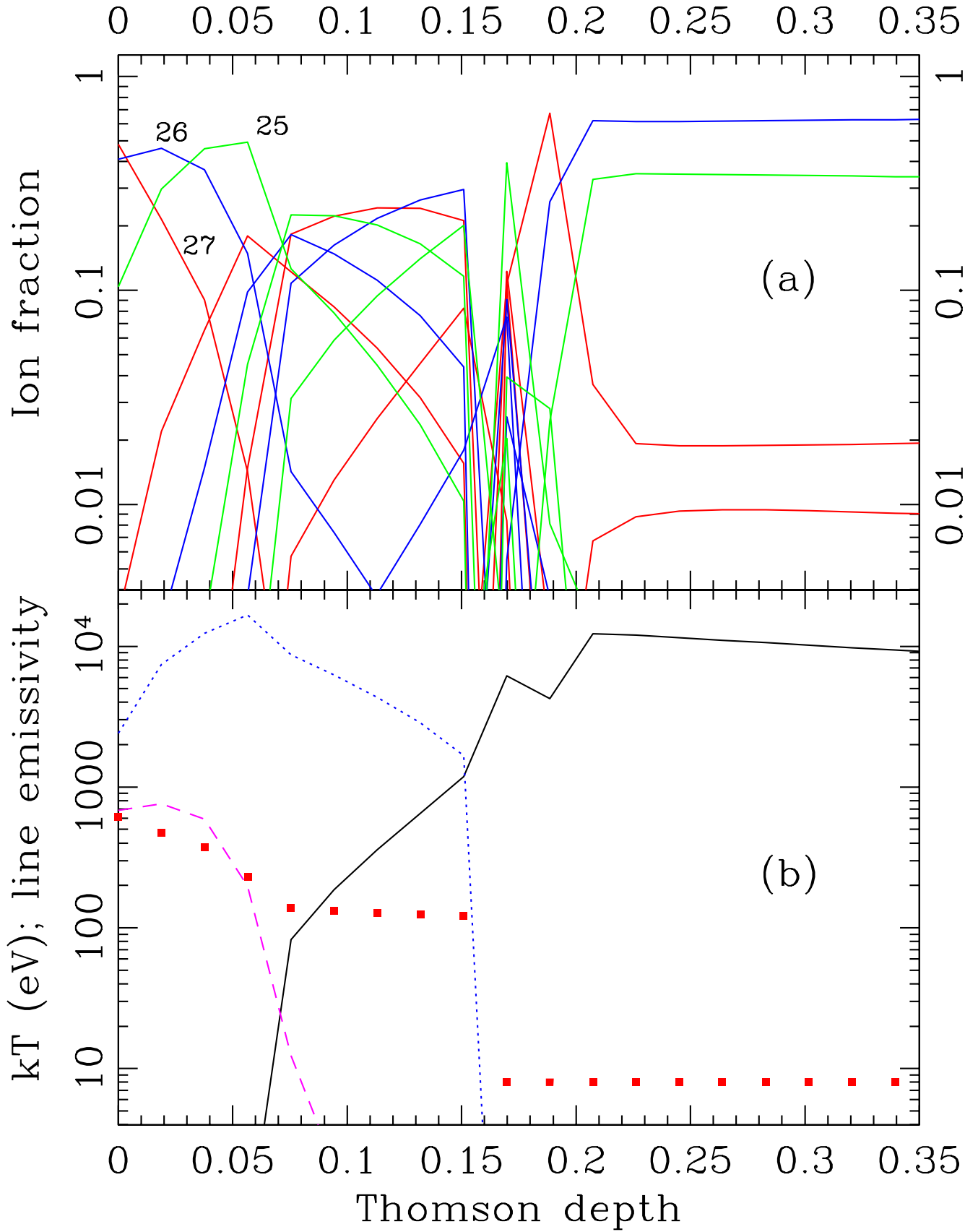}}
\caption{(a) -- Ionic fractions for iron as a function of Thomson
depth for the run $w3$. Note that the skin is far from being
completely ionized. (b) -- Line emissivity (arbitrary units) as a
function of Thomson depth for 6.4 keV, $\sim 6.7$ and 6.9 keV iron
line bins (solid, dotted and dashed curves respectively). The gas
temperature profile is also shown with filled boxes. Note that the
ionization structure and line emissivity change discontinuosly at the
temperature discontinuities.}
\label{fig:ils3}
\end{figure*}

The key to understanding the reflection spectra is the ionization
structure of the gas, shown in Figure (\ref{fig:ils3}) for iron for
run $w3$. Panel (a) of the Figure shows ionic fractions of different
ionization stages of iron (Fe27 $\equiv$ completely ionized iron) as a
function of the Thomson depth. These fractions are defined as the
ratio $n_i/n_{\rm Fe}$, where $n_i$ is the density of the ionization
stage $i$ and $n_{\rm Fe}$ is the total local density of iron. The
most important fact from the Figure is that the skin is highly
ionized, but it is far from being completely ionized. Even in
the uppermost zone, more than 50\% of iron is in the form of Fe26 and
Fe25, and this fraction quickly increases to almost 100\% at deeper
layers (and then goes to zero as less ionized ions start to
dominate). Panel (b) of Figure \ref{fig:ils3} shows the K$\alpha$ line
emissivity as a function of Thomson depth. The line emissivity is
split onto several components. In particular, the dashed line shows
the line emission in the energy bin that contains the H-like iron line
at $E\simeq 6.97$ keV; the dotted curve shows the line emission in the
bin that contains the He-like line at $E\simeq 6.7$ keV plus some
lines from intermediate ionization stages of ions Fe17-Fe23; and
finally, the solid curve shows the line emission in the bin that
includes the ``neutral''-like line from ions Fe1-Fe16 with $E\simeq
6.4$ keV.

Note that the skin emits a strong He-like line. Moreover, in the test
$w9$, whose ionization structure is shown in Figure \ref{fig:ils9},
the ``6.7 keV line'' bin dominates the line emissivity by far. In
addition, other strong features from the soft X-ray band to the iron
recombination continuum allow us to state the following general point:
because the warm skin is not completely ionized, it produces strong
atomic emission and absorption features that should be observable in
spectra of many bright AGN (if this model is the physically correct
one).

\section{X-rays from magnetic flares: hot skin limit}\label{sect:flares}

\subsection{Setup}\label{sect:flset}

We now will study the case in which the ionizing flux is much larger
than the disk soft flux, $\fx\gg \fdisk$, which is thought to be
appropriate for the two-phase patchy corona model of Seyfert Galaxies
(e.g., Haardt, Maraschi \& Ghisellini 1994 and Svensson 1996). In that
model, the X-rays are produced within magnetic flares (Galeev, Rosner
\& Vaiana 1979). The condition $\fx\gg \fdisk$ is a necessary
condition for the model to work, since it comes from the requirement
that Compton cooling of the X-ray {\em producing} active regions by
the soft disk photons not be too strong, because otherwise the
continuum spectra are too steep to explain typical Seyfert 1 spectra
and much less that of GBHCs in their hard state (e.g., Gierlinski et
al. 1997; Dove et al. 1997; Nayakshin 1998a).  As we discussed in NKK,
the X-rays induce evaporative winds close to the flare location, so
that hydrostatic balance does not apply in a direct sense. However,
based on earlier work on X-ray induced winds in stars (e.g., McCray \&
Hatchett 1975; Basko et al. 1977; London, McCray, \& Auer 1981, and
references therein) one expects to see the same two-layer structure
for the illuminated gas.  In the geometry of thin accretion disks, the
winds will decrease the value of the Thomson depth of the skin by
pushing the ionized material along the disk away from the flare
location. A careful multi-dimensional calculation including gas
dynamics, ionization and radition transfer is needed. Such calculation
is beyond the scope of this paper, but we attempt to model the effects
of the wind in a simple manner.

From the point of view of ionization calculations, the gas density
structure is of paramaunt importance for a self-consistent solution.
The gas density where the temperature discontinuity sets in is
expected to be the same or roughly the same in both static and wind
situations because this density is set by the energy balance that is
largely given by radiation heating/cooling and thus is independent of
whether the gas is static or moving non-relativistically. The density
gradient is then the quantity of the primary interest. In the
hydrostatic case, this gradient is given by the gravity in the
approximately isothermal skin (the gas temperature in the skin varies
by a factor of order of few only, see Nayakshin 2000). In the case of
the wind, the gas density gradient is controlled by the radiation
pressure, gravity and gas pressure forces. A simple estimate shows
that the ratio of the gravitational force to the radiation pressure
force is of the order of ${\cal F}_g/{\cal F}_{\rm rad}\sim
\fdisk/\fx$, which is much less than unity for magnetic flares. Hence,
the gas density gradient is controlled by the radiation pressure
mainly and is (very roughly!)  $\fx/\fdisk$ times the one which is
given by the gravity.

In the absense of a multi-dimensional radiation hydrodynamics and
radiation transfer approach, we resort to treating the complications
due to the wind by artificially increasing the value of the local
gravity in the atmosphere of the disk, which, in some sense, leads to
the same end result -- a lower value of $\tskin$ for a given $\fx$ (as
compared with the one that would have been obtained if there were no
winds). Mathematically this is done by multiplying the local gravity
gradient ($ -G M z/R^3$) by a dimensionless number ${\cal A}\geq 
1$. Note that this approach is similar to the one we adopted in most
of the calculations presented in NKK, although the gravity parameter
$A$ defined there is different from the parameter ${\cal A}$ defined
here.

In terms of the final results for the magnetic flare model, the poorly
constrained value of ${\cal A}$ means that the Thomson depth of the
ionized skin is {\em not} calculated exactly and represents a current
uncertainty of the model. However, we believe that, for a given value
of $\tskin$, the ionization structure of the illuminated gas is
calculated with a reasonble precision and thus this aspect of
calculations is reliable. Moreover, it is encouraging that the
uncertainties of the model due to the wind do not present an
unsurmountable obstacle because X-ray induced winds from accretion
disks and stars have been treated by many different authors (e.g.,
Buff \& McCray 1974; McCray \& Hatchett 1975; Begelman, McKee \&
Shields 1983; Proga, Stone \& Drew 1999). We plan to incorporate
evaporative winds in our calculations in the future.

\subsection{Temperature profiles}\label{sect:ftemp}

\begin{figure*}[T]
\centerline{\psfig{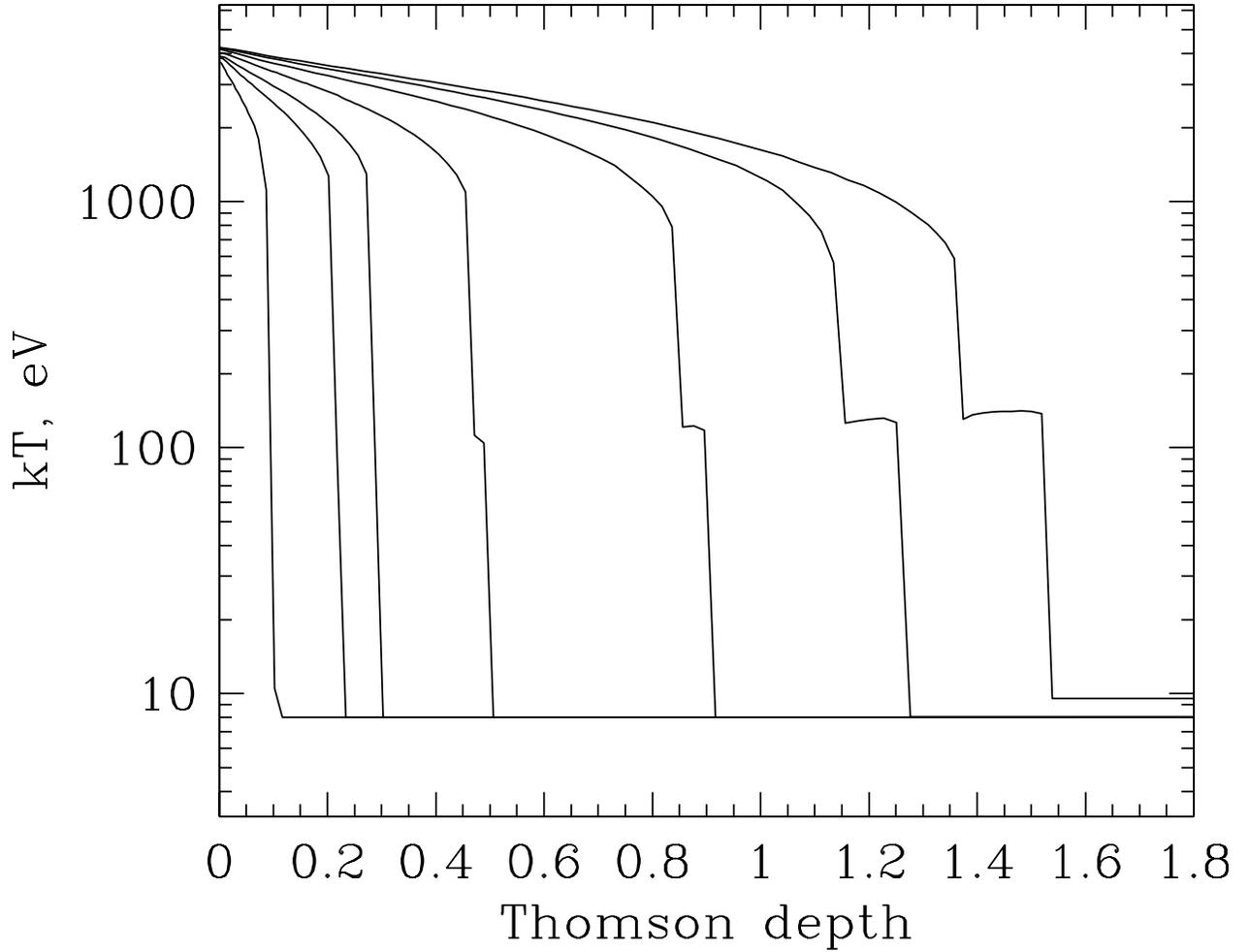}}
\caption{Temperature of the illuminated gas as a function of Thomson
depth for the ``strong illumination'' limit. The ratio $\fx/\fdisk$ is
fixed at a value of 75 for all the tests; radius $R = 6 R_S$; the
incidence angle is the same as in tests $w1--w9$ ($\pi/4$); the
dimensionless parameter ${\cal A}= 20$ for all the tests. The
accretion rate through the disk is varied from a low value of $\dm =
10^{-4}$ (test $h1$) to high of 0.0064 ($h7$). Note that with the
chosen value of $\fx/\fdisk$, the maximum value of the X-ray flux is
about that which would result from the lamppost if its luminosity were
about $\ledd$.}
\label{fig:temp1}
\end{figure*}

Figure \ref{fig:temp1} shows temperature profiles for the illuminated
gas. From these profiles, one can immediately see why we refer to the
limit $\fx\gg \fdisk$ as the hot skin limit. The Compton temperature
of the gas is much higher than that in the case $\fx\simlt \fdisk$
which can be understood from equation \ref{tc}. Although the {\em
X-ray} Compton temperature $T_x$ is the same in both cases, the local
Compton temperature is determined by the X-rays and the disk soft
flux, so that in the case $\fdisk\gg \fx$, $T_c$ is a small fraction
of $T_x$. Also note that the middle branch of the S-curve (the region
with $kT\sim 90- 150$ eV) does not appear in Figure \ref{fig:temp1}
until the skin becomes moderately thick. This is due to the fact that
the middle branch of the S-curve is available only when the average
photon energy in the overall illuminating specrum is low enough (see
\S \ref{sect:ie}). The latter condition can be satisfied in the given
case only by Compton downscattering of the incident hard X-rays, which
becomes significant when $\tskin\sim 1$.

\subsection{Reflected spectra}\label{sect:fsp}

\begin{figure*}[T]
\centerline{\psfig{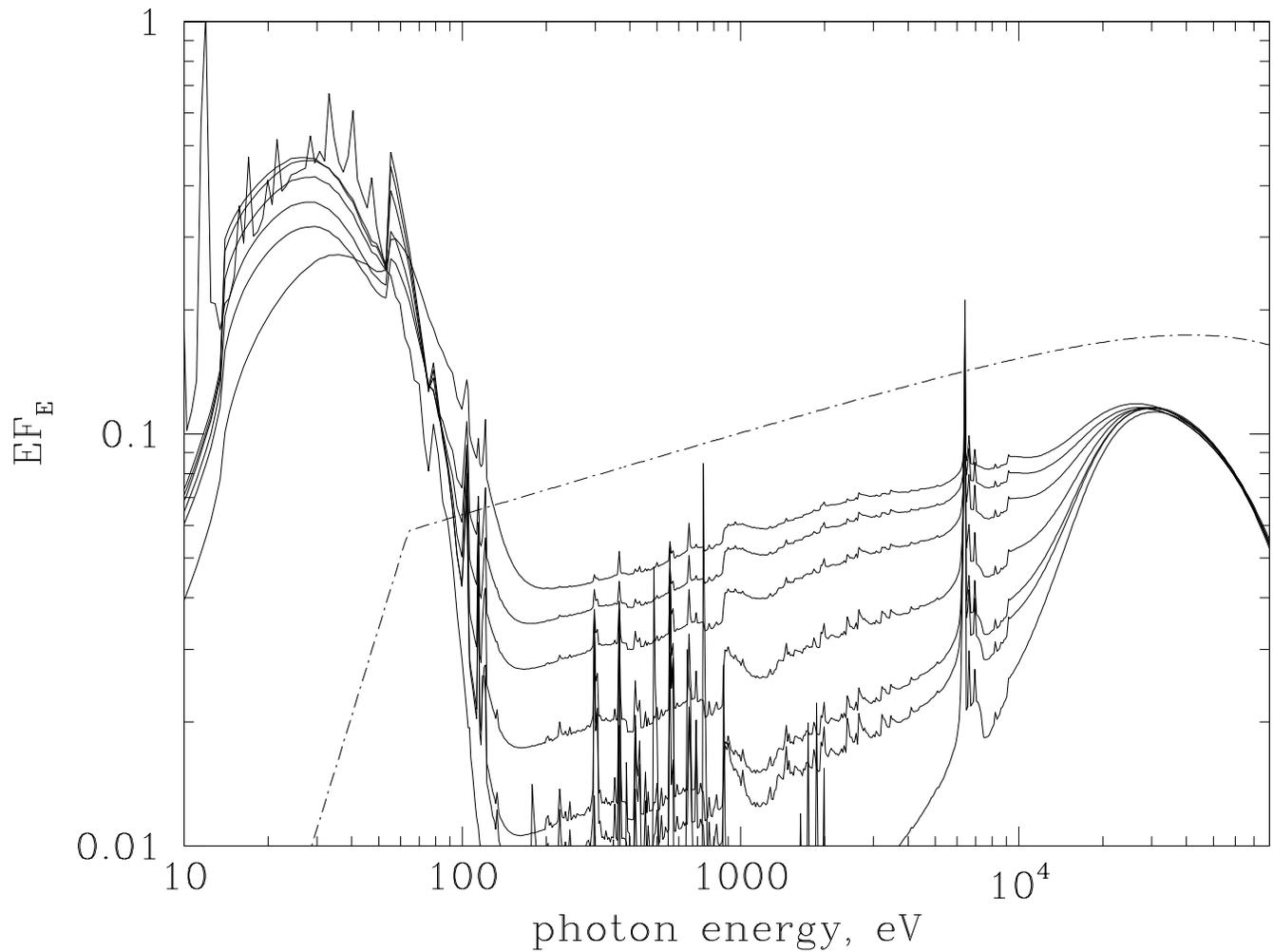}}
\caption{Angle-averaged reflected spectra for the runs shown in
Fig. \ref{fig:temp1}.  The most ionized case ($h7$) is the one with
the largest integrated X-ray albedo.  Note that the strongest line is
6.4 keV and that there is relatively little line emission in the soft
X-ray range.}
\label{fig:sp1}
\end{figure*}

Figures \ref{fig:sp1} and \ref{fig:sp2} show the reflected spectra for
the hot skin limit. This is the limit previously discussed by NKK. The
strongest line component is at 6.4 keV, the iron absorption edge is
neutral-like and is weak, and only few of the spectra exhibit some
soft X-ray emission. The latter is due to the fact that the skin is
more strongly ionized than it is in the warm skin limit, and thus
reflection in the soft X-ray range is dominated by the Compton
reflection in the skin. Below $\sim 30$ keV, the highest ionization
spectra can hardly be distinguished from the illuminating power-law,
especially if one adds the additional smearing due to relativistic
effects in vicinity of the black hole. This might be the reason why
GBHCs do not show broad iron lines and a somewhat small reflection
covering fraction (see Gierlinski et al. 1997 and references therein,
and also see Nayakshin 1998a; Ross, Fabian \& Young 1999; Done \&
Nayakshin 2000).

\subsection{Ionization Structure of the gas}

\begin{figure*}[T]
\centerline{\psfig{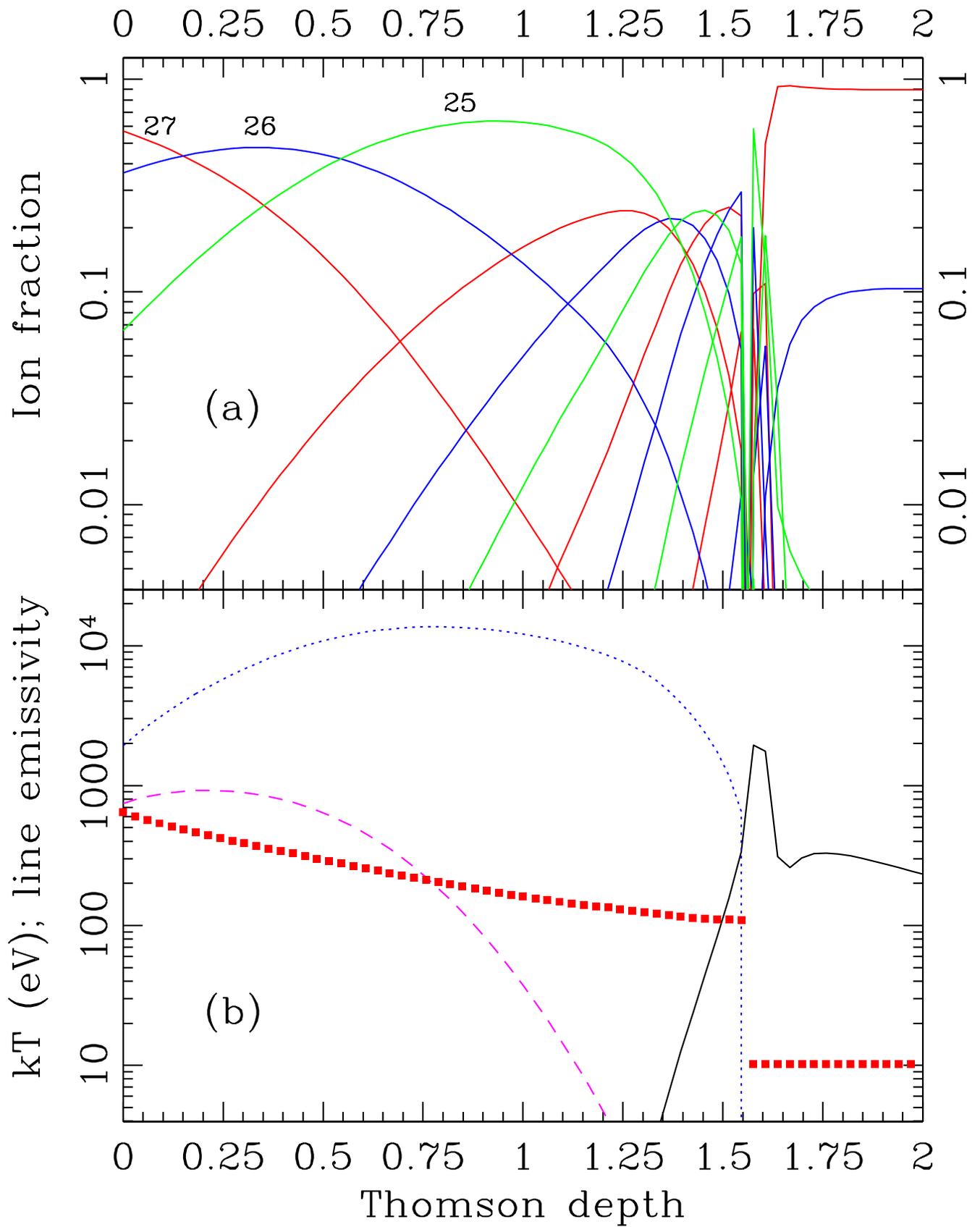}}
\caption{Same as Figure \ref{fig:ils3} but for test $w9$.}
\label{fig:ils9}
\end{figure*}

\begin{figure*}[T]
\centerline{\psfig{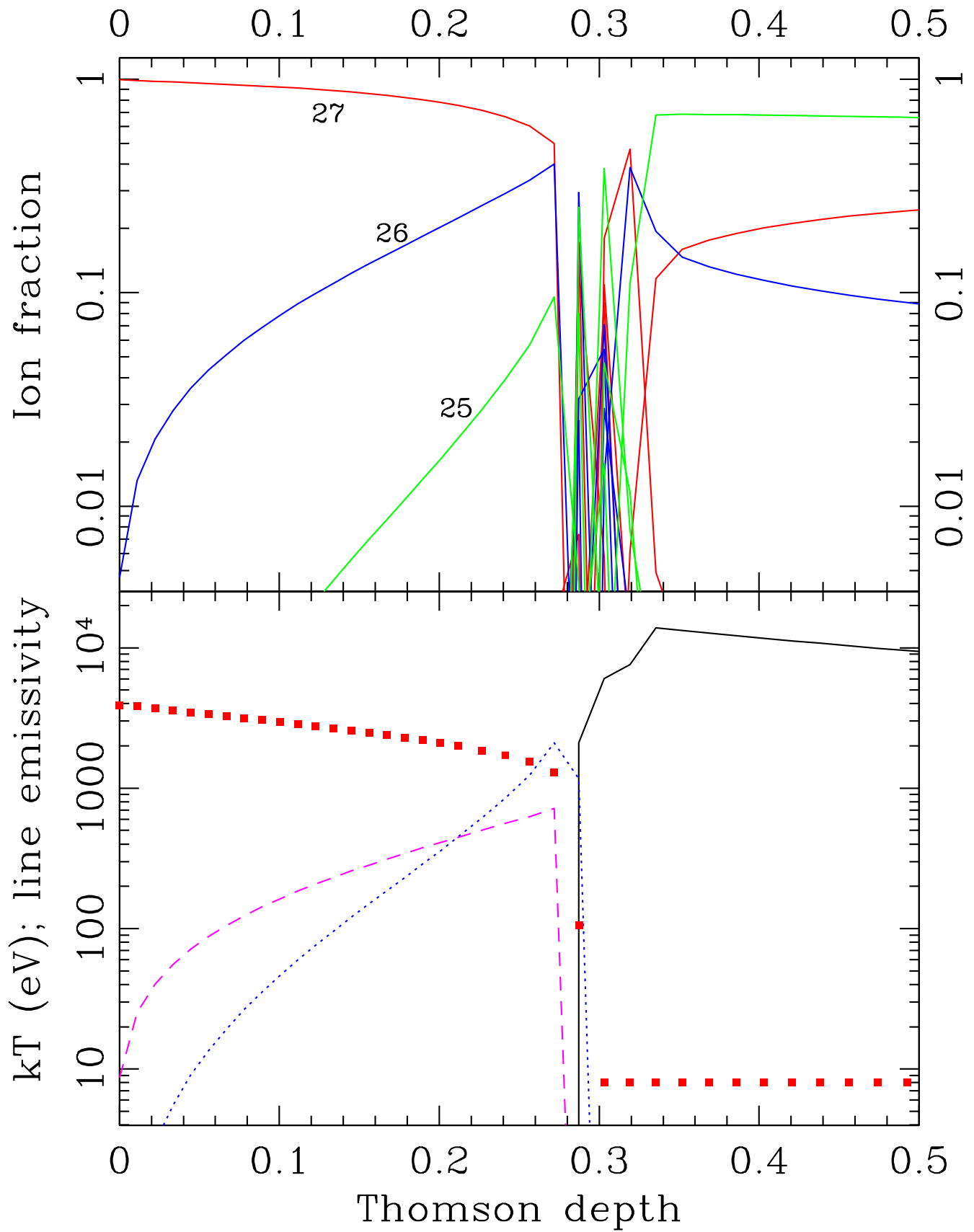}}
\caption{Same as Figure \ref{fig:ils3}, but for the hot skin run
$h3$. Note that the skin is almost completely ionized and emits weak
H-like and He-like lines. Their combined emissivity is much smaller
than that of the ``neutral''-like iron line (solid curve) below the
skin.}
\label{fig:il3}
\end{figure*}

\begin{figure*}[T]
\centerline{\psfig{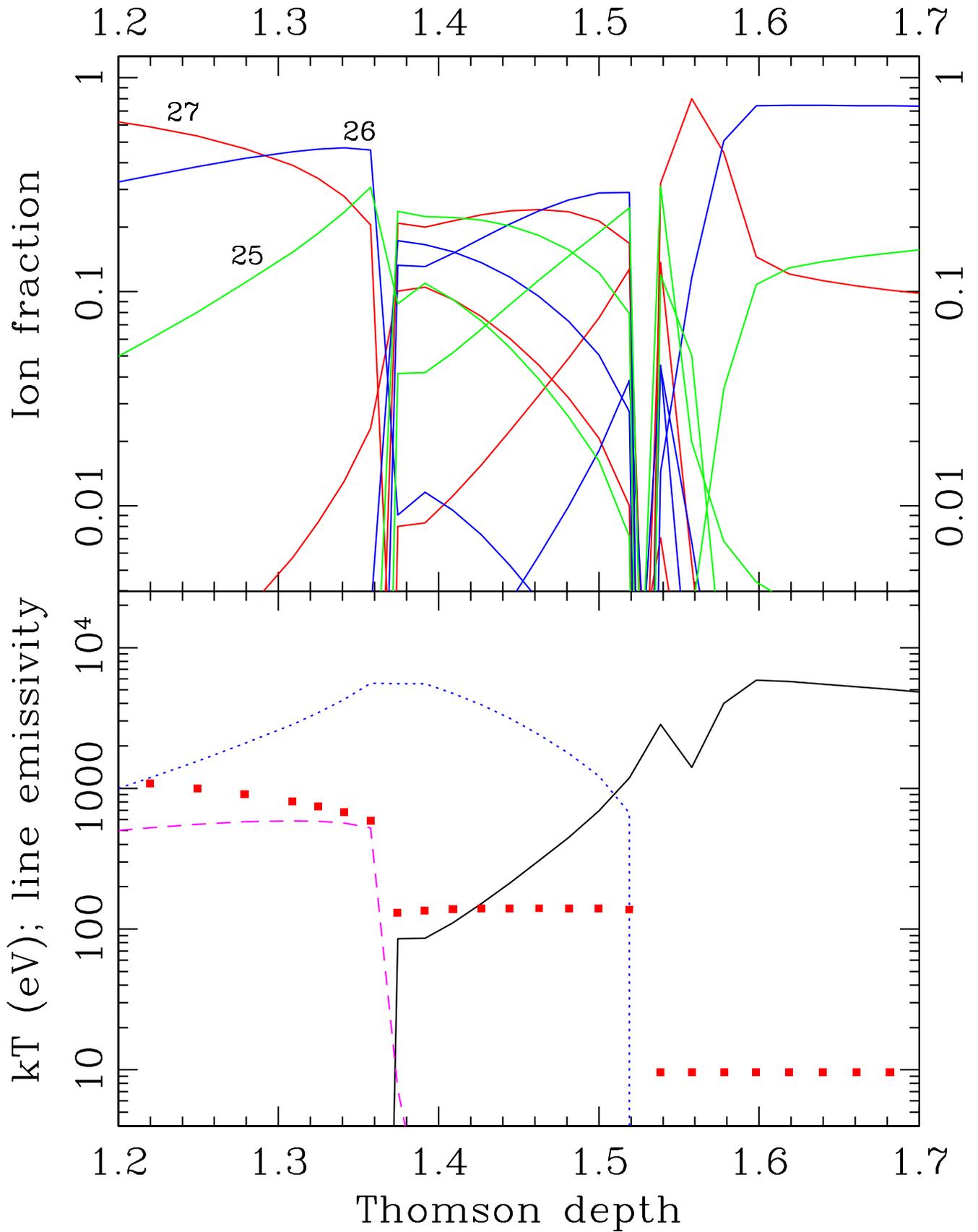}}
\caption{Same as Fig. \ref{fig:il3} but for run $h7$.}
\label{fig:il7}
\end{figure*}

Figures \ref{fig:il3} and \ref{fig:il7} show the ionic fractions and
line emissivity of the illuminated gas for tests $h3$ and $h7$,
respectively. These figures are to be compared with the figures
\ref{fig:ils3} and \ref{fig:ils9} for the warm skin tests $w3$ and
$w9$. The biggest difference is in the fact that the hot skin is
almost 100\% ionized except for regions near the temperature
discontinuity, whereas the warm skin is dominated by H- and He-like
iron. This is the reason why the line emissivity is dominated by the
``cold'' line coming from the cool material below the skin. The skin
thus only masks the presence of the cold gas. In addition, because the
skin is hot, Oxygen, Sulfur, Silicon and Neon are completely stripped
of their electrons, and thus there are no broad recombination edges or
line emission from these elements at least from the skin itself.

\section{Full corona above a standard disk}
\label{sect:fullcorona}

We will now consider X-ray reflection in the geometry of a full corona
covering the whole inner accretion disk. This geometrical arrangement
is physically rather different from the two other geometries that we
studied so far, because for the latter two, the X-ray producing region
does not directly border the illuminated surface of the disk. In other
words, there is no material above the top of the skin, and hence the
gas pressure at the top of the skin, $\pg(\ztop)$, is zero. This fact
is used as an explicit boundary condition and affects all aspects of
the X-ray illumination problem.  In the case of the full corona, the
pressure at the top of the skin is not zero, and by continuity
arguments, $\pg(z=\ztop) = P_{\rm cor}(\ztop)$, where $P_{\rm
cor}(\ztop)$ is the gas pressure at the bottom of the corona. Assuming
that the X-ray flux produced in the corona is radiated isotropically,
there is an equal amount of the X-rays emitted down to the disk and up
away from the disk.

Let us first assume that the corona is heated from below. In
equilibrium, there must be $F_{\epsilon} = 2 \fx$ {\rm heating} flux
from the cold disk to the corona. Some agent, most likely a magnetic
field with magnetic pressure $P_{\rm mag}$ will have to carry the
energy into the corona. The physical distinction between the skin and
the hot corona is then such that the magnetic fields heat the corona
but not the skin.  Further, $F_{\epsilon} \sim P_{\rm mag} v$, where
$v$ is the speed with which the energy flux is being carried. The
maximum value of $v$ is roughly the greater of $c_s$ and $v_A$, where
$c_s$ is the sound speed and $v_A$ is the Alfv\'en velocity.

Let the gas pressure at the top of the skin be related to the magnetic
pressure as $\pg = P_{\rm mag}/\beta$, where $\beta$ is a
dimensionless number. Hence, $v \sim c_s (1 + \beta^{1/2})$. The gas
pressure at the top of the skin is
\begin{equation}
\pg(\ztop) = \beta^{-1}\; P_{\rm mag} \sim {2 c\over v\beta}\, {\fx\over c}
\simeq {2 \times 10^{3}\; T_1^{-1/2}\over \beta (1
+\sqrt{\beta})}\, {\fx\over c}\;,
\label{ap1}
\end{equation}
where $T_1$ is the Compton temperature in units of 1 keV. It is
interesting to compare this pressure with the ``critical'' pressure
$P_c$. This quantity is defined to be the gas pressure at point (c) on
the S-curve (see Fig. \ref{fig:scurve}), i.e., at the location where
the temperature discontinuity occurs. Nayakshin (2000) found this
quantity to be
\begin{equation}
P_c = 0.032 \, T_1^{-3/2}\; {J_{\rm tot}\over c}\;,
\label{ap2}
\end{equation}
where $J_{\rm tot}\sim \fx$ is the total intensity of radiation
integrated over all angles at $z=\ztop$ (see also Krolik et al. 1981).
Thus,
\begin{equation}
{\pg\over P_c} \simeq 6 \times 10^{4}\; {1\over \beta (1
+\sqrt{\beta})}\; T_1^{2}\gg 1\;.
\label{ap3}
\end{equation}
That is, unless $\beta \equiv P_{\rm mag}/\pg \sim 10^3$ in the skin,
the gas pressure exceeds the one at which the hottest branch of the
solution (i.e., the completely ionized skin) exists. Note that
$\beta$, in fact, is likely to be less than unity. We defined the skin
as a region where no magnetic heating occurs. However, magnetic
reconnection often occurs when the magnetic pressure starts to exceed
the gas pressure, i.e., when $\beta \gg 1$.

This estimate was done for a corona heated from below. A different
scenario arizes when the corona is heated internally -- via viscous
dissipation of the accretion energy of the gas flowing through the
corona itself. In the latter case, the radiation flux from the corona
can be written as
\begin{equation}
\fx = \epsilon {3\over 8\pi} \,{GM \dot{M}_c\over R^3}\, J(R)\;,
\label{fcor}
\end{equation}
where $J(R) = (1- \sqrt{3R_s/R})$, and $\epsilon = 1$ for the complete
transfer of the accretion energy into radiation (as in Shakura-Sunyaev
disks), while $\epsilon$ is less than unity for an advection dominated
corona (e.g., Esin et al. 1997). Further, the accretion rate through
the corona, $\dot{M}_c = 2\pi R \Sigma_{\rm cor} v_{rc}$ where
$\Sigma_{\rm cor}$, and $v_{rc}$ are the corona mass column density
and radial inflow velocity, respectively.  The gas pressure in the
corona may be estimated via hydrostatic balance:
\begin{equation}
P_{\rm cor} = {1\over 2}\; {GMH_c\over R^3}\, \Sigma_{\rm cor}\;,
\label{pcor}
\end{equation}
where $H_c$ is the height scale of the corona. Using these two
equations, we can now conclude that
\begin{equation}
{P_{\rm cor} c\over \fx} = {2\over 3 \epsilon}\, {H_c\over R}\,
{c\over v_{rc}}
\label{prr}
\end{equation}
If the coronal accretion energy is radiated locally, then we can use
the standard equation for the radial inflow velocity: $v_{rc} \simeq
\alpha c_s (H_c/R)^2$, where $c_s$ is the sound speed in the corona.
Thus, one obtains
\begin{equation}
{\pg\over P_c} \simgt 21\, T_1^{3/2}\alpha^{-1}\;{c\over c_s}\;
{R\over H_c}\gg 1\;.
\label{prc}
\end{equation}
If coronal cooling is dominated by advection, then $\epsilon\ll 1$,
and so even though $H_c\sim R$ and $v_{rc}\sim v_K$, where $v_K$ is
Keplerian velocity (see, e.g., Narayan \& Yi 1994), $\pg$ is still
large compared with $P_c$.

Summarizing, in all of the three cases for the {\em full} corona above
the disk, we concluded that the gas pressure in the skin is very much
larger than the pressure at which the thermal instability
operates. Thus, the gas pressure and density below the corona are too
high for the Compton-bremsstrahlung stable branch of the solution to
exist, and hence {\em no ionized skin forms below the hot corona}. The
gas temperature below the corona is therefore close to the effective
one and the reflection and the lines will be those that are produced
in a ``neutral'' material. This conclusion holds for arbitrarily large
accretion rates. Note, however, that a thin transition layer may still
form due to conductive heating of the disk (e.g.,
Maciolek-Niedzwiecki, Krolik \& Zdziarski 1997).

\section{Discussion}\label{sect:discussion}

In this paper, we considered three physically distinct geometries: the
lamppost geometry; the magnetic flares above the disk and the full
corona above the disk. Each of these geometries was shown to produce
different reflected spectra. The large differences in the spectra were
caused by (i) absense or presence of a large soft disk flux that
influences the Compton temperature; and (ii) the additional pressure,
or weight, of the corona for the full corona geometry. While it is
perhaps possible to find conditions in which these three models may
yield similar spectra (one example is a very low accretion rate when
the ionized skin is very Thomson thin and thus negligible), the
behavior of the spectra with the X-ray luminosity as well as other
parameters is clearly different which should allow one to distinguish
between these models observationally.

We do not discuss observational status of the lamppost and the
magnetic flare models because we will present an extended discussion
of this topic in a future publication where we will also present
complete disk spectra that include relativistic broadening. However,
it appears to us that the full corona model is the most unpromising of
all and we will not study this model in our future work. In
particular, since no skin forms below the hot corona, it is not
possible to explain the hard continuum spectra of GBHCs (e.g., see
Gierlinski et al. 1997; Dove et al. 1997, Nayakshin 1998a). Moreover,
there are AGN that have more optical/UV power than the X-ray power
which argues against the full corona geometry (see Haardt et al 1994)
in which the reprocessed power should be rather less than the
X-rays. In addition, as we have shown in \S \ref{sect:fullcorona}, the
reflector below the skin is cold, which means that the small
reflection covering fraction seen in many GBHCs (e.g., Gierlinski et
al. 1997) is problematic. The lack of the iron line reverberation
discovered in the recent observations of two Seyfet 1 Galaxies (Chiang
et al. 2000; Reynolds 2000; Lee et al. 2000) presents yet another
uneasy challenge to this model.

Now we will concentrate our discussion on the differences between the
lamppost and the magnetic flare models.

\subsection{Changes in the S-curve due to soft disk flux}\label{sect:ie}

In this paper we emphasised the large difference in the resulting
temperature profiles, ionization structure and the reflected spectra
between the hot and the warm skin limits. For a clearer understanding
of our results, it is useful to discuss this difference in the
simplest case -- optically thin heating/cooling balance.  Figure
\ref{fig:scurve} shows the energy equilibrium curves for the spectral
index $\Gamma = 1.8$ and a constant angle-integrated intensity of
radiation $4\pi J = 10^{16}$ erg cm$^{-2}$ s$^{-1}$. The overall
intensity $J$ in these tests is a sum of the X-ray intensity, $J_x$
and the black-body intensity $J_{\rm bb}$ (with $kT = 10 eV$). The
only difference between the curves is the fraction of the black-body
intensity compared with that of the hard power-law. In particular, the
curves are computed for $J_{\rm bb}/J_x = $ 0, 1, 2, 4, and 8.

The temperature profiles shown in Figures \ref{fig:temps} and
\ref{fig:temp1} can now be discussed with the help of figure
\ref{fig:scurve}. The maximum temperature reached in the two opposite
limits is different due to the fact that the Compton-bremsstrahlung
stable branch of the S-curve shifts significantly as the ionizing
spectrum evolves from ``all X-ray'' to ``all black-body''.  The X-ray
Compton temperature for the given spectrum is about 7.1 keV. Stars in
figure \ref{fig:scurve} indicate the value of the Compton temperature
calculated from equation \ref{tc}, which is clearly smaller for larger
values of $J_{\rm bb}/J_x$. The fact that the corresponding curves
pass right through the stars shows that the gas temperature is indeed
very close to the corresponding Compton temperature when $\Sigma_x\gg
1$.

Another important difference between the temperature profiles for the
hot and warm skin limits is the extent of the middle stable shelf in
the S-curve. In the case of $\fx\simlt \fdisk$, the middle stable
shelf is always prominent in the temperature profiles, whereas in the
opposite limit when $\fx\gg \fdisk$, the middle stable branch only
appears for skin that is at least moderately Thomson thick. In order
to understand that, we recall that the thermal conduction picks
solutions with the smallest temperature gradients (see NKK), and thus
the transition from the Compton-bremsstrahlung branch to the middle
one happens after the gas pressure exceeds the pressure $P_c$ at point
$c$. Thus, in order for the middle shelf to be present in the
temperature profiles, the point $c$ should lie to the right of the
turning point at $kT\sim 80$ eV in Figure \ref{fig:scurve}.

When there is little blackbody flux, the gas pressure at point $c$ is
large enough that the transition happens directly from the
Compton-heated to the cold branch. As the blackbody intensity
increases, $T_c$ decreases, and then the transition happens first to
the middle branch and only later to the cold solution. However, with
increase in the blackbody intensity, point $c$ moves to the right as
much as to make the whole range of temperatures from $k T\sim 80 eV$
to the Compton one to be thermally stable.

In addition, even if $J_{\rm bb}=0$, the incident X-rays are
downscattered in the skin and thus for a moderately Thomson thick skin
point $c$ again moves to the right (the more the thicker the skin is).
This is best understood from the fact that the location of the point
$c$ sensitively depends on the Compton temperature, which is affected
by scatterings. Analytical theory of the Compton-bremsstrahlung cooled
branch (see Krolik et al 1981; Nayakshin 2000) yields that the gas
temperature at point $c$ is a third of the Compton temperature, and
pressure $P_c$ is given by equation \ref{ap2}.

The square boxes in Figure \ref{fig:scurve} indicate the location of
the critical point $c$ found via equation \ref{ap2}. Note that when
there is no black-body contribution to the ionizing intensity, the
analytical theory of the Compton-bremstrahlung cooled upper branch
works perfectly. Physically, this occurs because the gas is completely
ionized and atomic heating/cooling is negligible. However, when
$J_{\rm bb}\simgt J_x$, the gas becomes cool enough and heating due to
photo-absorption cannot be neglected; this is the reason why the
analytical solutions under-estimate the value of the gas temperature
at point $c$.

\subsection{Iron ionic fractions and lines}\label{sect:difion}

\begin{figure*}[T]
\centerline{\psfig{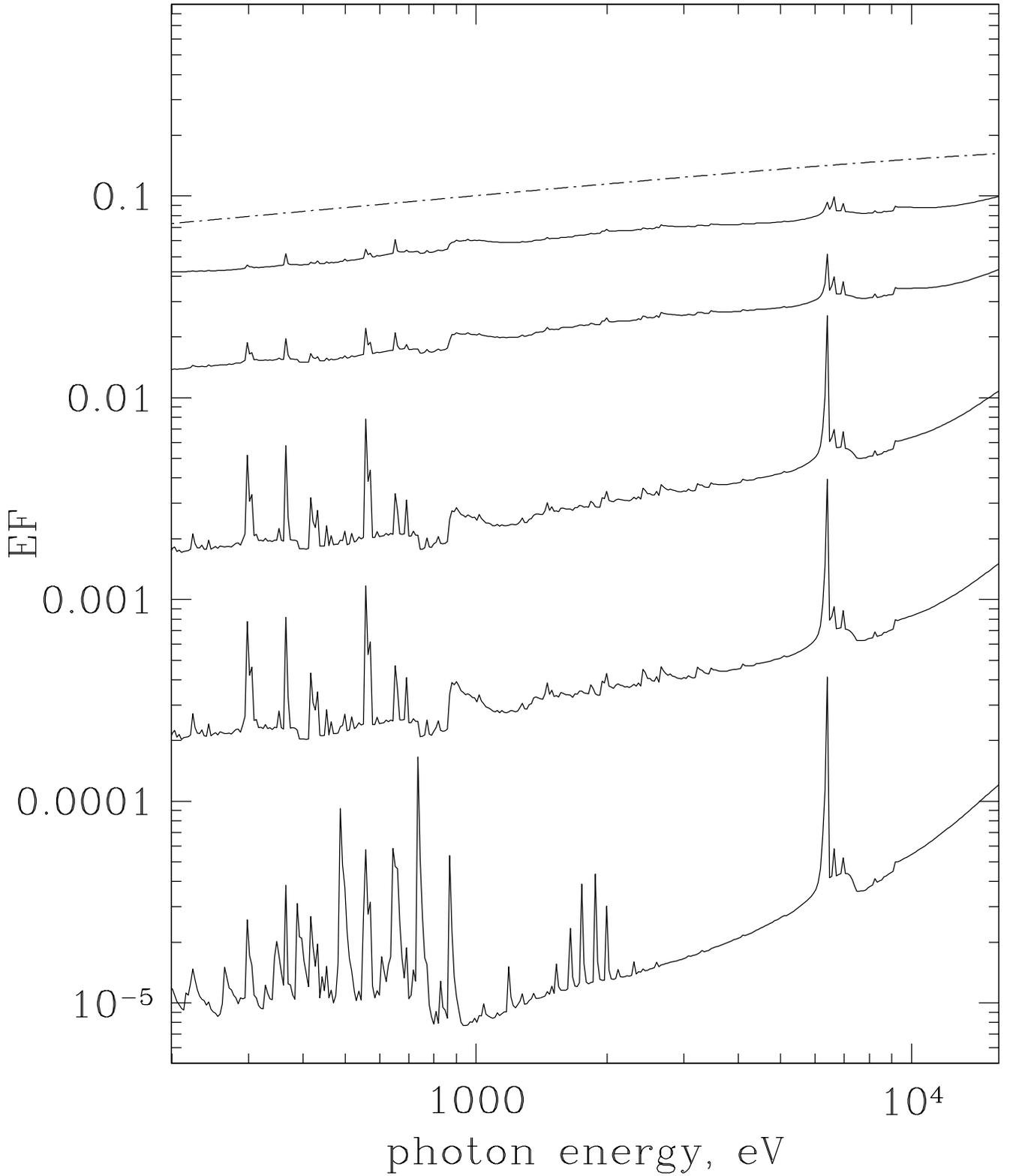}}
\caption{Angle-averaged reflected spectra for the runs $h1$, $h3$,
$h5$ and $h7$, zoomed in on the X-ray part of the spectrum. The curves
are offset with respect to each other to allow for visibility of the
line features. The dash-dotted curve shows the incident continuum. The
uppermost curve demonstrates that complitely ionized, mirror-like
reflection is the end point of the strong illumination in the case of
magnetic flares}
\label{fig:sp2}
\end{figure*}

\begin{figure*}[T]
\centerline{\psfig{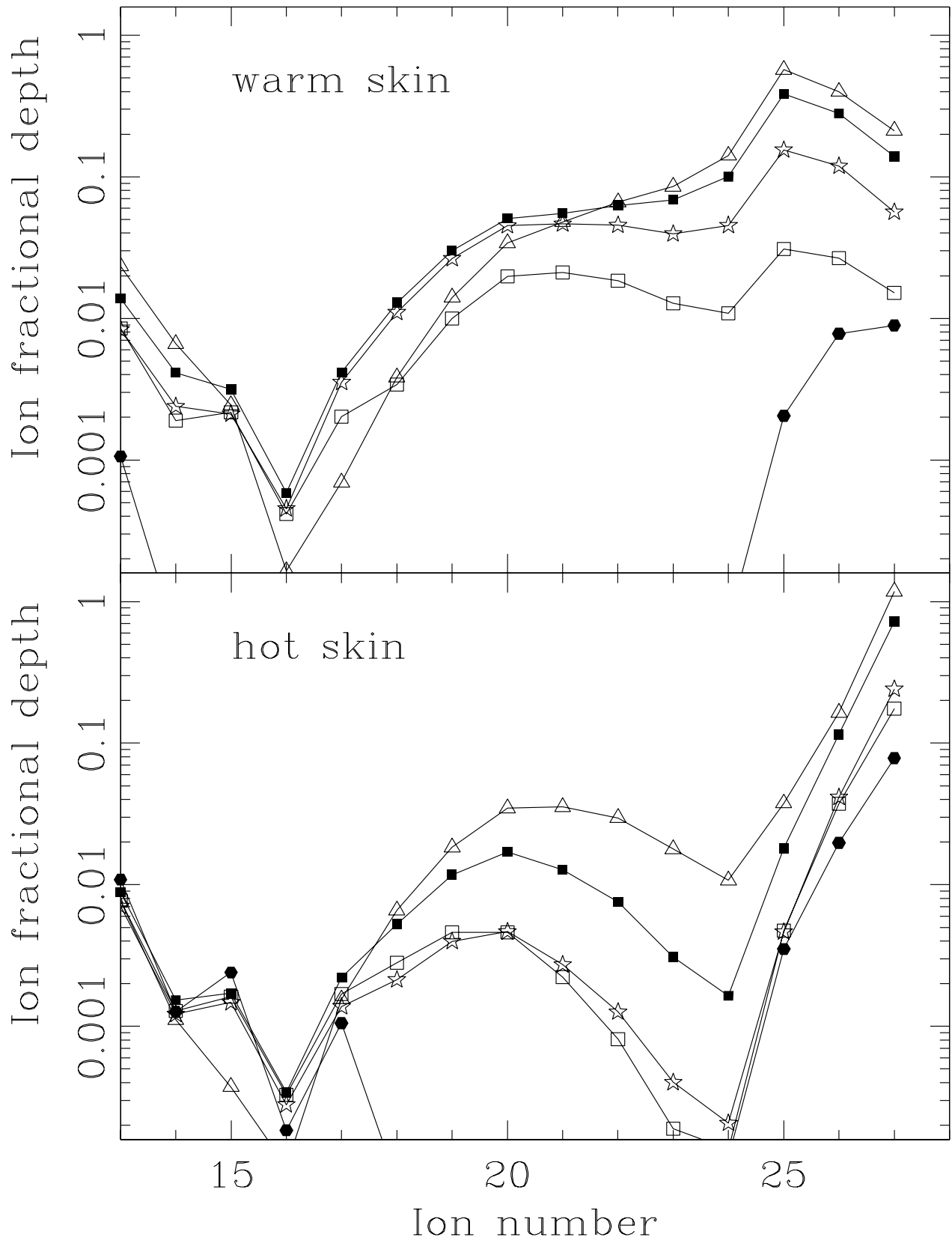}}
\caption{Integrated Thomson depth for different ionization stages of
iron for the warm skin limit (upper), and the hot skin limit (lower).
Note the predominance of completely ionized iron in the hot skin and
that of Helium-like iron in the warm skin, respectively.}
\label{fig:fractf}
\end{figure*}

Figure \ref{fig:fractf} summarizes the differences in the ionization
structure of the skin in the hot and warm limits by presenting the
integrated Thomoson depth of the ions Fe13 -- Fe27 for both of the
limiting cases. This depth is defined as the integral
$\int_0^{\infty}\, d\tau_t (n_i/n_{Fe})$, where $n_i$ is the ion
density for ion $i$. The curves shown in the Figure correspond to
tests presented in Figures \ref{fig:spssh} \& \ref{fig:sp2},
respectively. The largest difference between the ionization structure
of the two sets of calculations is seen for He- and H-like iron and
for completely ionized iron. For the warm skin cases, the first two
ions by far outweight the presence of Fe27, as well as any of the less
ionized stages. For the hot skin, the situation is reverse, and this
is why the skin is ``invisible'' to the observer in this case and can
only be uncovered in the hard X-ray energy range where relativistic
rollover in the Klein-Nishina cross section produce a corresponding
rollover in the reflected spectrum.


There is also a substantial difference in the equivalent width and
energy of the iron K$\alpha$ lines between the warm and hot skin
limits. In the former case, the most abundunt ion in the skin is the
He-like iron, and therefore the strongest line is also He-like with
energy around 6.7 keV. H-like line and lines from Fe17-23 also
contribute to the complex of the emitted lines. Because He-like iron
has a large fluorescence yield compared with neutral iron, the warm
skin actually yields the strongest iron line emission. In a sense, it
amplifies the line.

The opposite limit of the strong X-ray flux, $\fx\gg \fdisk$, is
characterized by the predominance of completely ionized iron in the
skin. The X-rays incident on the skin are Compton scattered but not
photo-absorbed with consecutive iron line fluorescence. Only those
photons that are able to reach the cold layers, whether scattered or
unscattered in the skin, will produce the ``neutral''-like K$\alpha$
line, which still needs to make it through the skin to the observer
(some of the line photons may be scattered in the skin, return to the
cold material and be photo-absorbed without re-emission as a 6.4 keV
line photon). Note also that scattering in the skin is very effective
in removing the photons from the line bin, because single scattering
disperses photons by $\Delta E/E \simeq 0.06 \, T_1^{1/2} \simgt 0.1$,
or about 700 eV at the line energy.  Therefore, the hot skin only
degrades the cold-like line by reflecting the incident X-rays and by
scattering the line photons emitted from the cool layers.

\subsection{On generality of our results}\label{sect:gen}

\subsubsection{Ratio of X-ray and disk fluxes}\label{sect:fxfd}

\begin{figure*}[T]
\centerline{\psfig{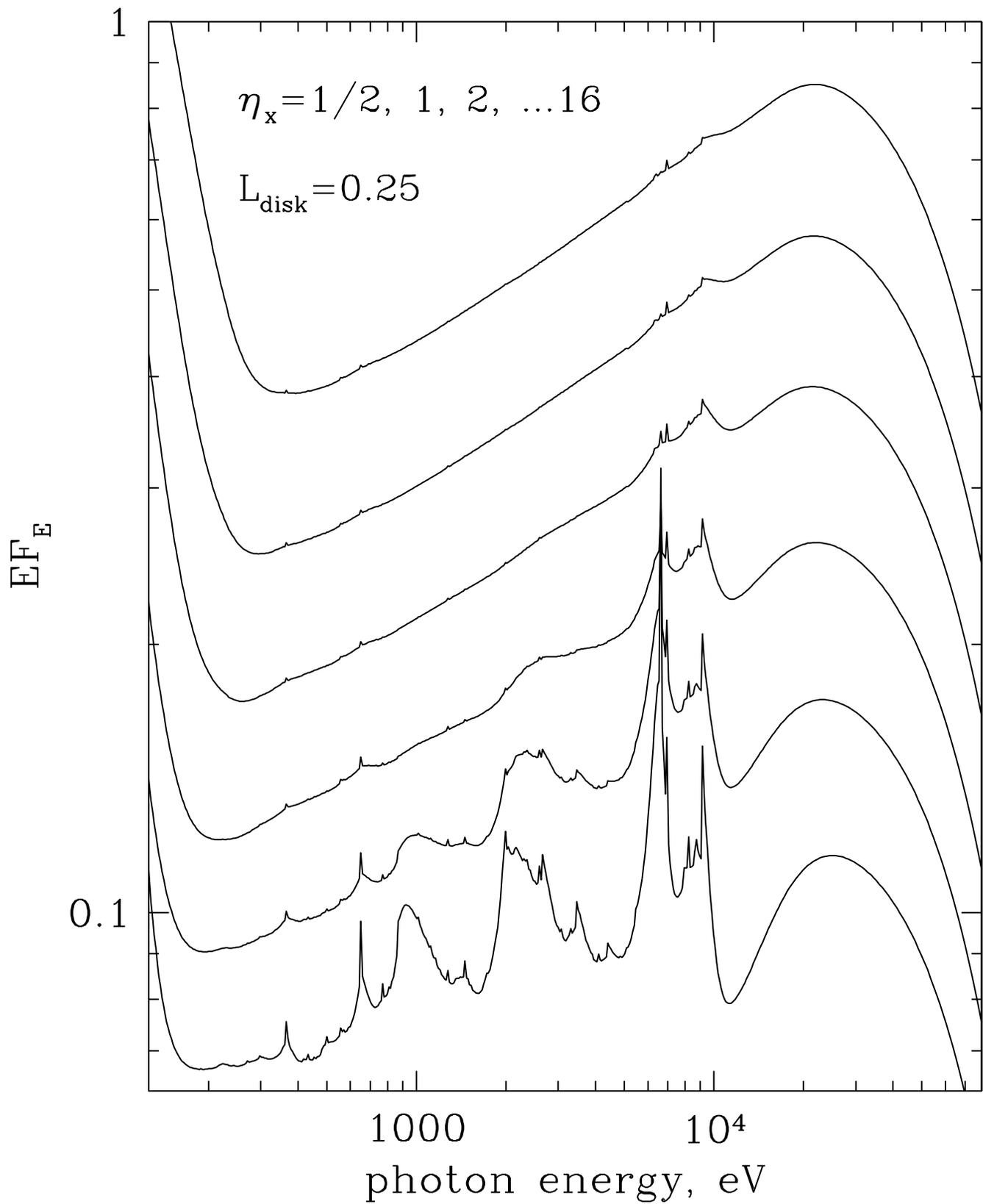}}
\caption{Reflected spectra for the lamp post model solid for a fixed
accretion rate through the disk, $\dm = L_d/\ledd =0.25$ and a
variable X-ray luminosity. From the bottom to top, $\eta_x = L_x/L_d =
2^{-1}, 1, ... 16$.}
\label{fig:vareta}
\end{figure*}

In terms of the ratio of the illuminating flux $\fx$ to the disk flux
$\fdisk$, we limited our attention so far to two rather extreme values
-- one quite small ($\eta_x = 0.2$) and the other very large,
$\fx/\fdisk = 75$. Thus, one wonders how our results will change when
$\eta_x$ is inbetween these two values. To address this question, we
conducted several additional series of tests. In the first series, we
fixed the accretion rate through the disk at $\dm = 0.25$, and varied
the ratio $\eta_x = L_x/L_d$ from $2^{-1}$ to $2^4$ in steps of factor
of 2. These tests are shown in Figure \ref{fig:vareta}. The reflected
spectra show a strong He-like line, edge and the recombination
continuum for all $\eta_x$ below 8. In fact, even for the largest
values of $\eta_x$ (8 and 16), there still exist a substantial column
density of H- and He-like iron, but it occurs rather deep in the skin,
i.e., at $\tau_T\sim 1$, so that it is smeared out by Compton
scatterings. The skin is very Thomson thick in these two latter tests
because $L_x$ is supper-Eddington.

\begin{figure*}[T]
\centerline{\psfig{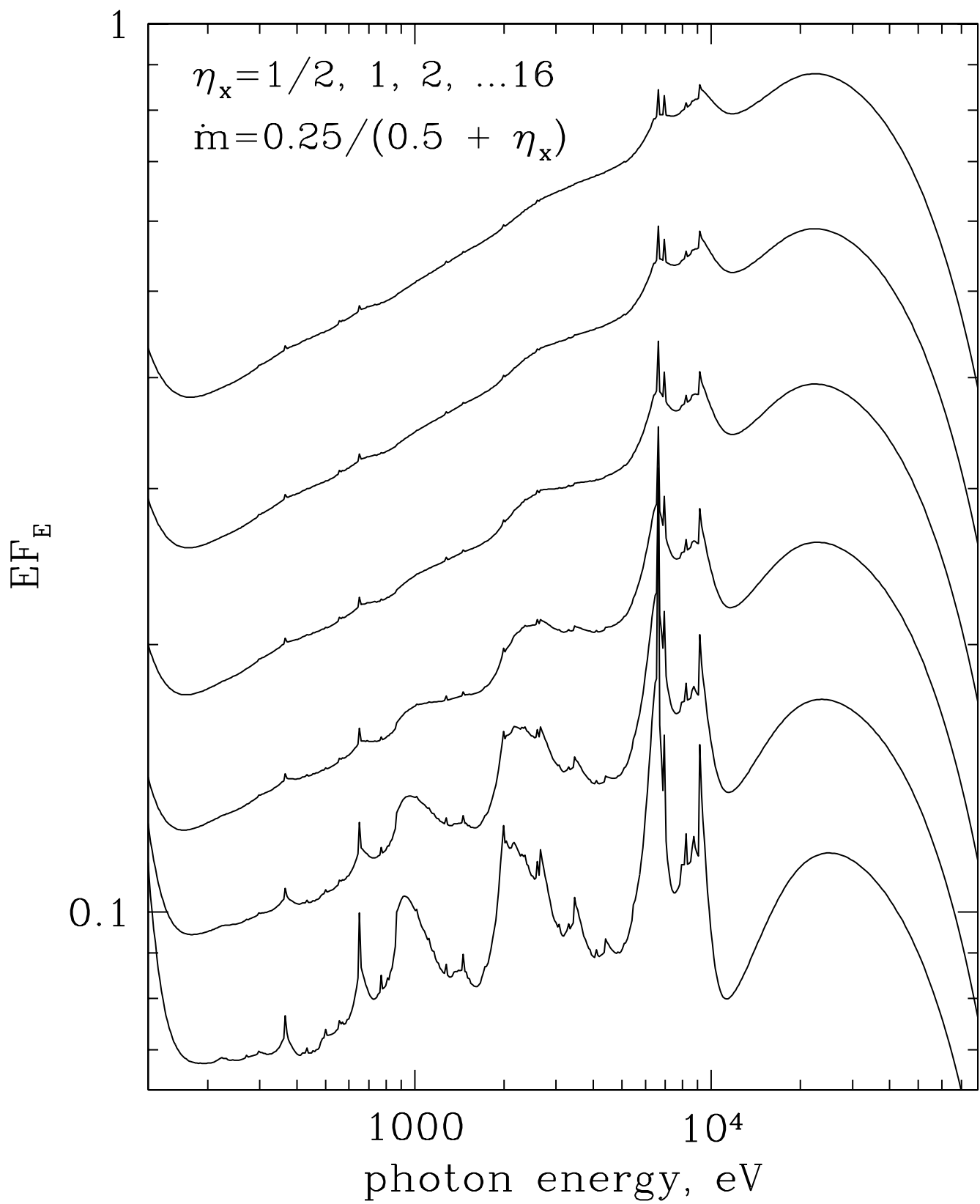}}
\caption{Same as Figure \ref{fig:vareta} but for the accretion rate as
shown in the figure.}
\label{fig:vareta1}
\end{figure*}

\begin{figure*}[T]
\centerline{\psfig{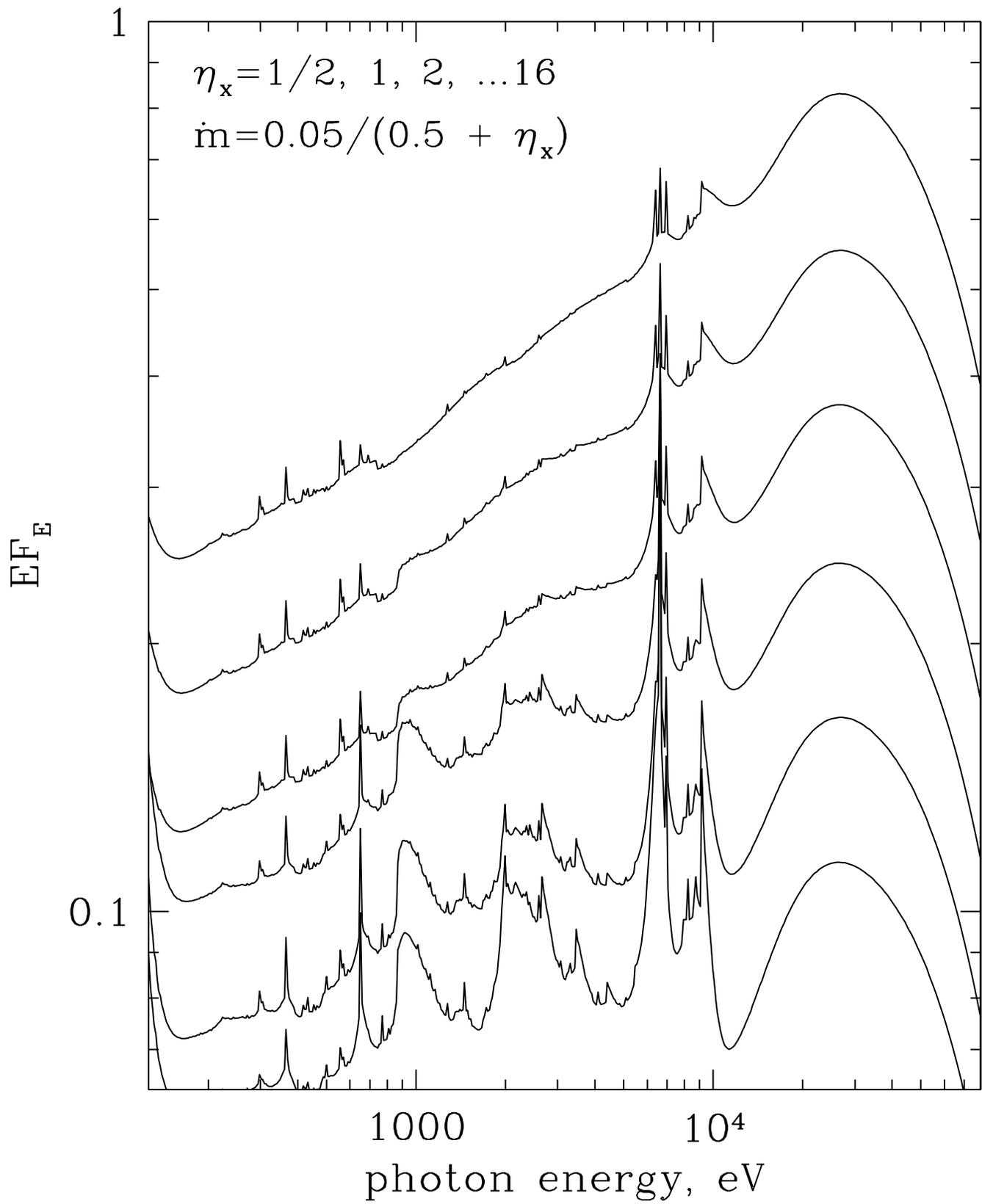}}
\caption{Same as Figure \ref{fig:vareta} but for the accretion rate as
shown in the figure.}
\label{fig:vareta2}
\end{figure*}

The fact that there is a large column density in the last two not
completely ionized stages of iron means that their presence may
probably be seen for lower values of $\dm$ when the completely ionized
part of the skin is not Thomson thick. To test that, we calculated
reflected spectra for same values of $\eta_x$, but for the accretion
rate through the disk varying as $\dm = 0.25/(0.5 + \eta_x)$. Results
of these tests are shown in Figure \ref{fig:vareta1}. Finally, we also
simulated reflected spectra for $\dm = 0.05/(0.5 + \eta_x)$, which we
show in Figure \ref{fig:vareta2}. Both of these series of calculations
indeed show that the Hydrogen and Helium-like iron still survives deep
inside the ionized skin even for $\eta_x = 16$. This is somewhat
unexpected since $\fx \simeq 10 \fdisk\gg \fdisk$ in the latter case,
and hence one could expect the skin to be in the hot limit. 

Let us try to explain this result. The gas temperature in the skin is
a fraction of the local Compton temperature, which is determined by
equation \ref{tc}. At large optical depth, one can use Eddington
approximation for the radiation transfer, and obtain that the
black-body radiation field is
\begin{equation}
J_{\rm bb} = \left\{ (1-a_s) + 
{F_{\rm disk}\over\fx}\right\}\, (1 + 3\tau/2)\;,
\label{eq9}
\end{equation}
where $a_s$ is the integrated X-ray albedo of the skin. When
$\tskin\gg 1$, the albedo becomes large, i.e., $1-a_s \ll 1$, and thus
in this situation one really compares $\fdisk/\fx$ with a small
quantity $1-a_s$ rather than with unity. Finally, when the Thomson
depth of the skin is greater than unity, Compton down-scattering of
the incident X-rays makes $T_x$ in equation \ref{tc} to be dependent
on the location in the skin -- it is lower on the bottom of the skin
than on its top. This is an additional mechanism to lower $T_c$. For
these reasons, it turns out to be enough for the disk flux to be a
small fraction of $\fx$ to present a large cooling source for the gas
{\em at the bottom of the skin}, and this is why one needs to go to
rather large values of $L_x/L_d\simgt 10$ to make the lamppost spectra
look similar to that from magnetic flares.

\subsubsection{The high energy rollover of the spectrum and spectral
index}\label{sect:roll}

Another degree of freedom which we have not explored in this paper is
the rollover energy, or equivalently, the gas temperature of the X-ray
{\em producing} region or regions. Our illuminating spectrum is chosen
to be reminiscent of what is typically seen in Seyfert 1 Galaxies and
GBHCs (e.g., Svensson 1996; Gierlinski et al. 1997), and this is why
we set the rollover energy at $E_c = 200$ keV. However, ``odd''
objects such as low-luminosity AGN do not have to have the same values
of $E_c$, and it appears to us to be potentially important for the
reflected spectra. In particular, $E_c$ figures prominently in the
determination of the X-ray Compton temperature $T_x$ if the X-ray
spectrum is hard (i.e., $\Gamma \simlt 2$). If $E_c$ is significantly
lower than 200 keV assumed in this paper, then the Compton temperature
of the skin may be low enough {\em even} if the spectrum is hard and
$\fx\gg \fdisk$. We will present concrete calculations showing these
effects elsewhere.

Finally, the spectral index $\Gamma$ also spans a range of values for
real sources, and it is well known that the reflected spectra strongly
depend on the actual value of $\Gamma$ (e.g., NKK). Notwithstanding
the complications connected with the two additional parameters, an
encouraging fact is that both of these parameters can in principle be
extracted from observations if a broad band spectrum is available.

\section{Summary}\label{sect:conclusions}

In this paper, we have sketched the way in which X-ray reflection
spectra may be computed for different accretion disk theories, and
then showed example calculations for two models (lamppost and flares)
and also discussed spectra from the full corona disk model. The main
three parameters that determine the outcome of the X-ray illumination
problem are (1) the ratio of the X-ray illuminating flux to the disk
thermal flux, $\fx/\fdisk$, because it defines the Compton temperature
of the corona and thus the degree to which it will be ionized; (2) the
accretion rate through the disk since it defines the height at which
the skin is located ($H$ to few $H$); (3) Compton temperature that is
a function of the spectral index of the ionizing radiation, $\Gamma$,
and the spectral cutoff energy, $E_c$. These parameters can be readily
deduced from observations once UV and X-ray spectra, luminosities and
at least an estimate of the black hole mass for the given source are
available. Therefore, for any accretion disk model which clearly
specifies a connection between the overall observed X-ray luminosity
$L_x$ and the radius-dependent ionizing flux $\fx$, accurate
disk-integrated spectra can be calculated. These spectra are weakly
dependent on the unknown value of the viscosity parameter $\alpha$
(see \S 2).

\begin{figure*}[T]
\centerline{\psfig{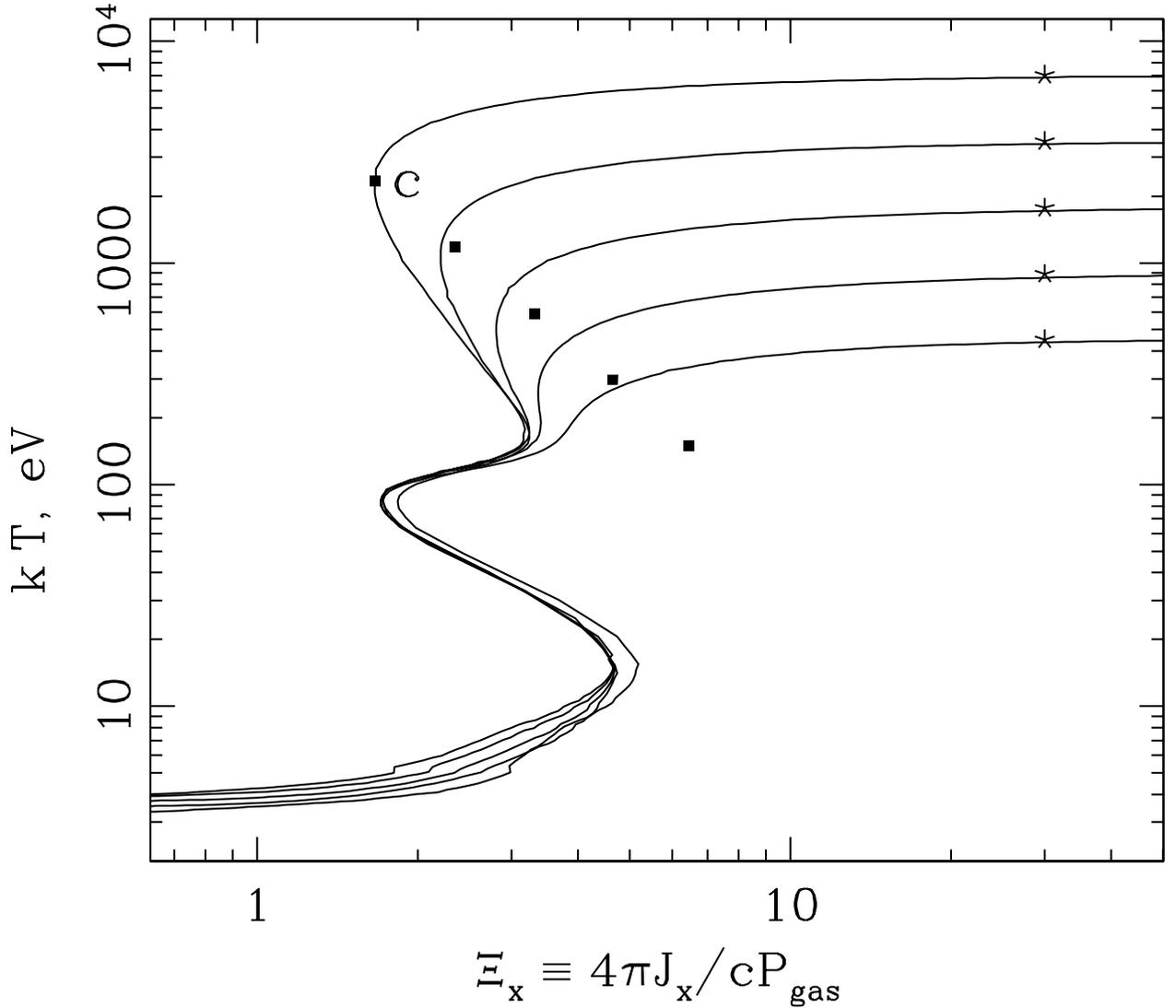}}
\caption{Ionization equilibrium curves (S-curves) computed for an
optically thin (both in continuum and lines) gas illuminated by a
$\Gamma=1.8$ ionizing continuum (same spectral form as for all the
tests in this paper) and a blackbody with $kT = 10 eV$. The total
angle integrated intensity, $4\pi J = 4\pi (J_x + J_{\rm bb})$ is kept
constant at $10^{16}$ erg cm$^{-2}$ s$^{-1}$, and the fraction of the
blackbody is varied: $J_{\rm bb}/J_x = 0$, 1, 2, 4, 8, from the top to
the bottom curves, respectively. The stars in the figure show the
Compton temperature calculated with equation \ref{tc} and filled
squares give the approximate analytical prediction for the gas
temperature and pressure at the discontinuity (see equation
\ref{ap2}).}
\label{fig:scurve}
\end{figure*}

Our main results, valid for relatively hard X-ray spectra ($\Gamma
\simlt 2$) and the rollover energy $E_c \sim 200$ keV, are as
following:

\begin{itemize}
\item 
    If the incident X-ray flux is smaller than or comparable with the
    soft thermal flux generated intrinsically in the disk, then the
    Compton-heated skin is ``warm'', i.e., $kT\simlt 1$ keV. In that
    case iron in the skin is not completely ionized and the majority
    of it is in the form of hydrogen and helium-like ions. In this
    limit, the iron line is dominated by the He-like line at 6.7 keV
    and the line equivalent width increases with the ionizing X-ray
    luminosity.

\item
    In addition, in the warm skin limit, the ionization physics
    permits the existence of a large shelf of material occupying the
    middle stable branch of the S-curve (see Figure
    \ref{fig:scurve}). Medium Z-elements such as oxygen are not
    completely ionized there and thus they produce strong soft X-ray
    features which should be visible with modern X-ray telescopes such
    as Chandra and XMM.

\item
    If the illuminating X-ray flux is much higher than the soft disk
    flux, then the skin temperature is substantially higher. Iron is
    then mostly completely ionized in the skin and thus the skin does
    not emit or absorbs photons due to atomic processes. Therefore,
    the hot skin only masks the presence of the cold material. The
    larger the ionizing X-ray luminosity, the less atomic marks one
    sees in the reflected spectra. In particular, the iron line is
    emitted almost exclusively by the cold material below the skin
    and its energy is close to 6.4 keV. Further, EW of the line
    decreases with increasing $L_x$ and eventually goes to zero.

\item
    Because ionization equilibria depend on the ratio $\fx/\fdisk$,
    the middle temperature solutions are not present in the hot skin,
    and thus there are less soft X-ray emission features from medium
    Z-elements. For high $L/\ledd$ in the magnetic flare model, the
    reflected spectra are featurless power-laws up to the Compton
    rollover at $\sim 30$ keV.

\item
    Full coronae above accretion disks do not have the highly ionized
    skin between the corona and the disk, because the gas density and
    pressure are too high to allow for that. This means that X-ray
    reflection from full corona models, if not broadened to
    invisibility by scattering in the corona, is always ``cold'' and
    does not change with increase in the accretion rate through the
    disk or the corona. We believe this model is rulled out
    observationally (see \S \ref{sect:discussion}).
\end{itemize}

We hope that the significant differences in the spectra of these three
models make it possible to distinguish among them observationally
based on current and future data.

The authors acknowledge many stimulating discussions with Manuel
Bautista, Demos Kazanas, and Julian Krolik. SN acknowledges National
Research Council Associateship which fully supported this research.

{}

\end{document}